\documentclass[twocolumn, longauthor, twocolappendix]{aastex701}

\usepackage{siunitx}
\shorttitle{Aquila Clusters}
\shortauthors{Fielder et al.}

\newcommand{\tnm}[1]{\tablenotemark{#1}}

\defcitealias{Fielder2024}{F24}

\begin{document}

\title{
Fragmentation in the Serpens/Aquila Star-forming Region
}

\author[orcid=0009-0001-8625-505X]{Samuel D. Fielder}
\affiliation{Department of Physics and Astronomy, University of Victoria, Victoria, BC V8P 1A1, Canada}
\email[show]{samuelfielder@uvic.ca} 

\author[0000-0002-5779-8549]{Helen Kirk}
\affiliation{Herzberg Astronomy and Astrophysics Research Centre, National Research Council of Canada, 5071 West Saanich Road, Victoria, BC, V9E 2E7, Canada}
\affiliation{Department of Physics and Astronomy, University of Victoria, Victoria, BC, V8P 1A1, Canada}
\email{helenkirkastro@gmail.com}

\author[0000-0003-0749-9505]{Michael M. Dunham}
\affiliation{Department of Physics, Middlebury College, Middlebury, VT 05753, USA}
\email{mdunham@middlebury.edu}

\author[0000-0003-1252-9916]{Stella S. R. Offner}
\affiliation{Department of Astronomy, The University of Texas at Austin, Austin, TX 78712, USA}
\email{soffner@utexas.edu}

\begin{abstract}
We present a population study of Atacama Large Millimeter/submillimeter Array (ALMA) Cycle 6 observations of the 100 most gravitationally unstable dense cores in Aquila using a simple mass versus size analysis. We identify 66 continuum sources from ALMA 12~m observations at 106~GHz and through comparisons with known protostellar catalogs; two of these detected dense cores appear to be completely starless, without any accompanying/nearby protostar detections. Additionally, we find nine other starless ALMA 12~m detections within protostellar cores that have fragmented into a mixture of starless and protostellar substructures. We test the turbulent core collapse model by conducting synthetic observations of turbulent magnetohydrodynamical simulations of collapsing starless cores in order to predict how many starless cores should be detected given their central density and density profile. The simulations predict at least one (1.19) detection, consistent with our two detections of ALMA 12~m emission within completely starless cores. We also use a combination of ALMA Compact Array Cycle 4 observations and the Herschel Gould Belt Survey data to analyze how mass is distributed on three distinct spatial scales, in order to understand how turbulence shapes the evolution of substructure development as dense cores collapse to form new star systems. We find an increase in multiplicity at the smallest scales when the parent larger-scale structure also has a higher degree of fragmentation.
\end{abstract}


\section{Introduction}
\label{sec:intro}

Dense cores are subparsec density-enhanced regions typically found in filamentary structures within molecular clouds, which host the beginnings of future star systems \citep{DiFrancesco2007, Andre2014}. Understanding how dense cores collapse to form protostars is still a fundamental question that bears further exploration \citep{McKee2007, Pineda2023}. Due to the large number of physical processes like magnetic fields, turbulence, self-gravity, thermal support, feedback mechanisms, which span multiple order of magnitude of spatial scales, the picture is still not clear on which processes dominate in the low-mass star forming picture.

Over the past few decades, turbulence has emerged as one of the principal physical mechanisms responsible for the generation of overdensities within molecular clouds at small scales ($\sim$0.1~pc). Additionally, in a dual role, turbulence can act as a regulating mechanism to counter self-gravity at large global scales \citep[see][and the references therein]{Chevance2023}.

There exist several families of models describing the stability and evolution of dense cores. In some, cores evolve in a relatively slow and quasi-static fashion, gradually adjusting their structure while resisting collapse for several freefall timescales \citep{Larson1969, Shu1977}. In others, supersonic turbulence induces local compression, which leads to cores shaped by highly dynamic, turbulent processes that generate inhomogeneities \citep[e.g.,][]{MacLow1999, Hartmann2001}, where only a fraction of these overdensities becomes gravitationally bound and proceed toward eventual protostellar collapse \citep{Offner2022}. From this perspective, the longevity of a core depends on the interplay between stabilizing influences, such as turbulence, magnetic fields, and external pressure, and destabilizing ones that tip the balance toward collapse. The full range of models thus represents a continuum between long-lived, marginally stable structures and short-lived, rapidly collapsing ones, with environmental conditions playing a decisive role in determining whether a core can remain stable and ultimately form stars.

The advent of high-resolution radio facilities like the Atacama Large Millimeter/submillimeter Array (ALMA) has significantly transformed our understanding of the core scale ($<1000$~au), particularly with respect to the emergence and characterization of core substructure \citep[e.g.,][]{Tokuda2014, Zhang2015, Kong2017}. Observatories like ALMA are able to systematically inspect large dense core populations across the nearby ($<500$~pc) star forming regions with minimal observing time. For example, the ALMA Survey of Orion Planck Galactic Cold Clumps (ALMASOP) observed a total of 72 young dense cores in the Orion region, chosen to statistically investigate the evolutionary stage and substructures within \citep{Dutta2020}. If a starless core is gravitationally collapsing and reaches sufficient density, it will contain a high-density region that is detected as compact substructure by (interferometers like) ALMA. The detection of this substructure indicates that the core is prestellar and will likely soon become protostellar. If there are multiple structures/protostars in an individual core, this indicates a forming multiple star-system. All collapsing cores are expected to develop high-density compact regions \citep{Offner2012}, and many groups have been searching for this with ALMA \citep[e.g.,][]{Dunham2016, Kirk2017-Oph, Fielder2024}, including the ALMASOP group, who have identified high-density compact substructures within prestellar cores in the Orion complex \citep[e.g.,][]{Dutta2020, Sahu2021, Sahu2023, Hirano2024}. Additionally, owing to its proximity, the Taurus molecular cloud complex has been extensively investigated as a benchmark region for low-mass star formation \citep{Hacar2011,Hacar2013,Palmeirim2013,Tokuda2016}. \citet{Tokuda2014} presented the first ALMA observations of the dense core MC27/L1521F, revealing intricate substructures on sub-1000~au scales and indicating that the earliest stages of core evolution may be governed by highly dynamical initial conditions.

Our group has been involved in a series of population studies of dense cores using ALMA in the star forming regions of Chameleon I \citep{Dunham2016}, Ophiuchus \citep{Kirk2017-Oph}, and Orion B North \citep[][hereafter \citetalias{Fielder2024}]{Fielder2024}; this paper is a continuation of those efforts. The aforementioned Ophiuchus and Orion B North studies (unlike the Chamaeleon I study) both found evidence of substructure within dense cores, showing that the number of detections reflected an agreement with turbulent core collapse models. The difference in local environment may have a profound effect on which physical mechanisms contribute most to the collapse of dense cores \citep{Offner2022, Moon2025}. While Ophiuchus and Orion B North have ongoing star formation, dense cores in Chamaeleon I may be dispersing rather than collapsing to form protostars, leading to the lack of detectable substructure in the starless core population \citep{Belloche2011, Dunham2016}. 

While the physical processes governing core stability and collapse establish the initial conditions for star formation, an important consequence of collapse is fragmentation. Large, unbiased surveys of dense cores capable of systematically characterizing fragmentation remain rare. \citet{Pokhrel2018} conducted a comprehensive fragmentation analysis across a hierarchical range of scales, from the larger cloud structure ($\sim$10~pc) down to the protostars ($\sim$10~au). At all scales probed, \citet{Pokhrel2018}'s observed number of substructures was significantly lower than what classical thermal Jeans analyses would predict; this deficit in fragmentation could be understood if additional physical mechanisms are included, including turbulence. Analyses of larger cluster-forming high-mass regions show good agreement with thermal Jeans fragmentation \citep[e.g., ][and references therein]{Ishihara2024}, while Polaris and Lupus, diffuse regions that may be forming few, if any, new stars, are perhaps different \citep{Ishihara2025}.

Aquila/Serpens is an area of active star formation with fragmentation seen across all scales \citep{Pokhrel2023}.
The southern portion of the Aquila Rift region is host to the main star forming areas of Serpens South, W40, and MWC297 \citep[for an illustration, see Figure 1 of][]{Pokhrel2023}. Serpens South is a protostar-rich young cluster forming along a filamentary structure first discovered by \citet{Gutermuth2008}, while W40 is a nearby HII region that has cleared most of its molecular gas content \citep{Rumble2016}. Additionally, to the south of the two aforementioned regions is MWC297, a smaller region named after the highly reddened O-type star at its center \citep{Hamaguchi1999}.

A wide range in distances has been reported for Aquila Rift clouds over the years. Values range from $\sim$260~pc \citep[e.g.,][]{Straizys1996} to upwards of $\sim$650~pc \citep[see][and the references therein]{Zhang1988}. Many early studies were indirect measurements using extinction from dust grains, pointing to a distance of 260~pc where the front wall of the extinct feature of the Aquila Rift cloud complex was detected \citep[e.g.,][]{Straizys2003}. Newer very long baseline interferometry measurements put the constituents of the Serpens South at a distance of $\sim$436~pc \citep{Ortiz-Leon2018}, suggesting that there may be a large zone of low extinction material between us and the main cloud. Further, \citet{Ortiz-Leon2023} showed conclusively through astrometry measurements of H$_2$O masers in the Serpens South core and W40 complex, a distance of $436.0\pm9.2$~pc to the young protostars. Many other studies also confirm that the star forming regions may be placed farther than the front of the extinction wall \citep[e.g.,][]{Kolesnik1983, Shuping2012}. Therefore, we adopt a distance measurement of $d=436$~pc for this paper.

We present ALMA 3~mm continuum data for the 100 most gravitationally unstable cores in the Aquila Rift South region, as determined using the HGBS observations \citep{Konyves2015}. In Section \ref{sec:observations}, we discuss the dense cores targeted and the ALMA main array Band 3 data, while we outline our detections and any protostellar associations in Section \ref{sec:detections}. In Section \ref{sec:derived_properties}, we derive estimates for the mass and number densities of our detections, followed by brief discussions on the peak and integrated fluxes of our detections. In Section \ref{sec:simulations-synthetic-observations}, we use simulations of turbulently collapsing cores to predict the number of expected detections in our dataset. In Section \ref{sec:multi-scale-analysis}, we incorporate our ALMA Compact Array (ACA) 7~m observations, publicly available JCMT Gould Belt Survey SCUBA-2 450~$\mu$m observations, and the original Herschel observations to put into context the amount of fragmentation across different spatial scales. We summarize the findings in Section \ref{sec:conclusions}.

\section{Observations}
\label{sec:observations}

\subsection{Target Selection: HGBS}
\label{sec:target-selection}

The Herschel Gould Belt Survey \citep[HGBS;][]{Konyves2015} created a census of all dense cores identified in the southern Aquila cloud complex covering the extended region surrounding Serpens South and W40, as well as the more southern MWC297. Using observations from both the PACS and SPIRE instruments, along with core identifying algorithm \texttt{getsources} \citep{Menshchikov2012}, a total of 651 starless cores were identified along with 58 protostellar sources \citep{Konyves2015}. Of the identified starless cores, approximately 60\% appear to be gravitationally bound (named prestellar cores) determined by comparisons to the thermal value of the critical Bonnor-Ebert (BE) mass. Using the estimated dust temperature from an SED fit of the Herschel data, the derived core mass and observed core radius were used to identify the most 100 unstable cores for follow-up detailed observations, consisting of 31 protostellar sources and 69 prestellar sources (see Section \ref{sec:protostellar-reclassification} for details on our updated protostellar classification).

\begin{figure*}
    \centering
    \includegraphics[]{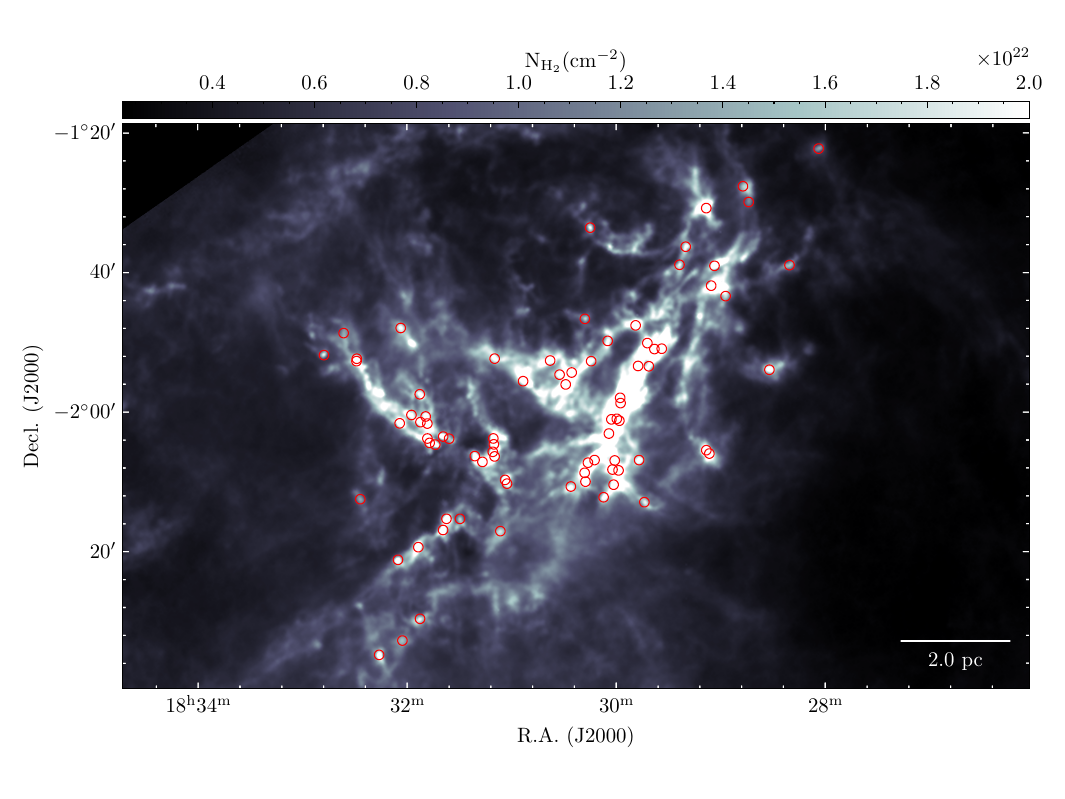}
    \caption{
    Column density map of the Serpens South portion of the Aquila Region at \ang{;;18.2} resolution, from \citet{Konyves2015}. The red circles represent a portion of the 100 dense cores observed by ALMA, with the diameters of the circles representative of the 12~m primary beam. We overplot a scale bar, computed with an adopted distance to Aquila of 436~pc.
    \label{fig:hgbs-pointings-overplotted}
    }
\end{figure*}

As outlined in Section \ref{sec:intro}, we adopt a distance of $d=436$~pc for this study; however, the original HGBS study assumed a distance of $d=260$~pc \citep{Konyves2015}. We therefore update the HGBS core mass, core size, and number density estimates given the increase in assumed distance, $d_{\text{rat}}=(436~\text{pc}/260~\text{pc})$.
The core masses scale as $d_{\text{rat}}^2$ and increase by a factor of approximately 2.8, the core sizes scale as $d_{\text{rat}}$ and increase by a factor of approximately 1.7, and the number density estimates therefore scale as $d_{\text{rat}}^{-1}$ and decrease by approximately 0.6. For detailed discussion on the differences in the original HGBS core catalog for these two choices in distances, see Appendix C in \citet{Konyves2015}.

\subsection{ALMA Observations}
\label{sec:alma-observations}

\subsubsection{ALMA 12~m Band 3 Data}
\label{sec:alma-12m-band3-data}

The ALMA 12~m Band 3 data (\#2018.1.00197.S, hereafter 12~m data) were observed between 2018 December 18 and 2018 December 22, and consisted of three unique observation times, using an average of 42 antennas (baselines 15-500~m). Four spectral windows were used, with three configured for continuum measurements, centered at 101, 103, and 113~GHz, each with a bandwidth of 2.00~GHz. Additionally, the last spectral window was configured for observation of $^{12}$CO$(1-0)$ emission at 115~GHz with a bandwidth of 0.059~GHz. We use the line-free spectral channels of this last spectral window along with the three other basebands to construct the continuum maps, while we use the line emission to detect areas of protostellar outflows as needed (see Section \ref{sec:alma-co-data} for details). Flux and bandpass calibrators were observed at the beginning of each execution block, followed by single pointings of all 100 science targets, with periodic phase and gain calibrators therein. The continuum observations were requested to have a 1$\sigma$ rms noise of 0.1~mJy~beam$^{-1}$, which was achieved for most 12~m pointings. The average synthesized beam is $\ang{;;1.3}\times\ang{;;1}$ ($\sim$570~au $\times$ 440~au) at a position angle of 75$^{\circ}$, along with a maximum recoverable scale of approximately \ang{;;18}.

\subsubsection{ALMA-ACA 7~m Band 3 Data}
\label{sec:alma-7m-band3-data}

The ALMA-ACA 7~m Band 3 data (\#2016.1.00320.S, hereafter 7~m data) were observed between 2017 March 21 and 2017 July 26, and consisted of six unique observation times, using an average of nine antennas (baselines $9-45$~m). The baseband configuration is the same as the 12~m observations (see Section \ref{sec:alma-12m-band3-data}). Due to the natural inclusion of larger-scale structures in the 7~m data, along with less collecting power of the smaller array, the achieved rms noise is larger than the 12~m dataset, ranging from approximately $1.00-1.49$~mJy~beam$^{-1}$. The averaged synthesized beam is $\ang{;;14}\times\ang{;;10}$ ($\sim$ 6100~au $\times$ 4360~au) at a position angle of 80$^{\circ}$, along with a maximum recoverable scale of approximately \ang{;;66}.

\subsubsection{Calibration and Reduction}
\label{sec:calibration-reduction}

For our dataset, the 12~m array's uv-plane coverage is sufficiently dense at short spacings that the 7~m data adds negligible information when combining the 12~m and 7~m datasets during imaging (combined imaging). With the combined images showing no additional large-scale emission recovery and peak fluxes agreeing to within 1\%, we have chosen to treat the two datasets independently in the calibration and reduction procedures described below.\footnote{The seldom negligible improvement from combination imaging (as in our case) is actively being investigated; see \citet{Petry2024} for more details.}

The 12~m data calibration and reduction were conducted with the pipeline v5.6.1 of the Common Astronomy Software Applications \citep[\texttt{CASA};][]{TheCasaTeam2022} software. Additionally, the 12~m data was passed through the \texttt{auto-selfcal}\footnote{\url{https://github.com/jjtobin/auto_selfcal}} software version 1.3, which conducts an automated self-calibration procedure on the continuum to improve image quality. This is only successful on observations where enough signal-to-noise is present, which in our case, was SNR$\approx$31. The attempted iterative solutions conducted had phase solution interval lengths of \texttt{inf}, 12.10~s, and \texttt{int}, with a final amplitude calibration with a solution interval of \texttt{inf}. A total of 18 fields were successfully self-calibrated with varying solution interval lengths: three fields for \texttt{inf}, six fields for 12.10~s, five fields for \texttt{int}, and four fields for fields amplitude calibrated.

For imaging and analysis of the 12~m dataset, \texttt{CASA} v6.5.0 was utilized for its most up-to-date auto-masking routines \citep{TheCasaTeam2022}. To construct all continuum images, the line emission was first subtracted from the last spectral window in order to make use of all available continuum emission, along with standard choices of imaging parameters: Briggs weighting with a robust value of $R=0.5$, along with default auto-masking parameters for the 12~m data.

The 7~m data calibration and reduction were conducted with the pipeline v4.7.2 of \texttt{CASA}, and no automated self-calibration was conducted due to the relatively lower SNR level achieved for the dataset. For imaging and analysis of the 7~m dataset, \texttt{CASA} v6.5.0 was also utilized for its most up-to-date auto-masking routines \citep{TheCasaTeam2022}; we reduced both the \texttt{sidelobethreshold} parameter to 0.75 (from the default 1.25) and \texttt{lownoisethreshold} parameter to 1.0 (from the default 2.0), to better capture larger-scale emission surrounding brighter areas of emission. This has the effect of more easily extending the mask region used for cleaning in the major iterations of the CLEAN algorithm \citep{TheCasaTeam2022}, allowing deeper cleaning to be conducted on areas with low but significant emission.

For individuals fields that overlapped in coverage with other individual fields, mosaicking was used to improve the final sensitivity of the image. There are a total of 41 ALMA 12~m fields and 58 ACA 7~m fields that are found in mosaics. In the mosaicked images, the only change made to the standard imaging parameters is the use of the \texttt{mosaic} gridder option.

In Table \ref{tab:observation-noise-levels}, we provide the information of the observed ALMA pointings, including their formal field names and positions on the sky. Additionally, we provide the name of the mosaicked field if the individual field overlapped with any other field. We also show the mosaicked field information for the analogous 7~m pointings. Finally we give the rms noise of the field, as well as the protostellar classification.

\begin{deluxetable*}{lllccccc}
\tabletypesize{\scriptsize}
\digitalasset
\tablecaption{
Noise Levels of Targeted ALMA Observations \label{tab:observation-noise-levels}
}

\tablehead{
    \colhead{Field Name} &
    \colhead{R.A.} &
    \colhead{Decl.} &
    \colhead{12m Mosaic\tnm{a}} &
    \colhead{12m $1\sigma$ rms\tnm{b}} &
    \colhead{7m Mosaic\tnm{a}} &
    \colhead{7m $1\sigma$ rms\tnm{b}} &
    \colhead{YSO?\tnm{c}} \\
    \colhead{} &
    \colhead{(J2000)} &
    \colhead{(J2000)} &
    \colhead{Field Name} &
    \colhead{($\text{mJy}~\text{beam}^{-1}$)} &
    \colhead{Field Name} &
    \colhead{($\text{mJy}~\text{beam}^{-1}$)} & 
    \colhead{}
}

\startdata
182513.1-025955 & 18h25m13.14s & -02d59m55.00s & ... & 0.08 & ... & 1.00 & Y \\
182730.2-035032 & 18h27m30.27s & -03d50m32.50s & ... & 0.08 & ... & 1.09 & N \\
182754.3-034237 & 18h27m54.36s & -03d42m37.20s & ... & 0.09 & 182754.3-034237M & 1.04 & Y \\
182757.5-034046 & 18h27m57.52s & -03d40m46.00s & ... & 0.08 & 182754.3-034237M & 1.04 & N \\
182758.8-034948 & 18h27m58.85s & -03d49m48.30s & ... & 0.08 & ... & 1.15 & N \\
182803.8-012214 & 18h28m03.89s & -01d22m14.80s & ... & 0.08 & ... & 1.06 & N \\
182809.2-034809 & 18h28m09.22s & -03d48m09.90s & ... & 0.17 & ... & 1.41 & Y \\
182820.5-013857 & 18h28m20.58s & -01d38m57.00s & ... & 0.08 & ... & 0.96 & N \\
182830.9-034733 & 18h28m30.93s & -03d47m33.40s & ... & 0.08 & ... & 1.00 & N \\
182832.0-015356 & 18h28m32.09s & -01d53m56.70s & ... & 0.09 & ... & 1.10 & N \\
... & ... & ... & ... & ... & ... & ... & ... \\
\enddata

\tablecomments{The full version of this table is available in machine-readable format in the online journal. A portion is shown here for guidance regarding its form and function.}

\tablenotetext{a}{For individual fields that overlap in coverage, the mosaic field name is given as the eastern-most field ending in ``M."}

\tablenotetext{b}{1$\sigma$ rms noise, computed in nondetection (unmasked) areas of the field. This value was computed from the nonprimary beam corrected image.}

\tablenotetext{c}{Protostellar classification primarily based on the \citet{Konyves2015} dataset. Sources marked ``Y-R" are those labeled as prestellar in \cite{Konyves2015}, which we reclassified as protostellar based on more recent catalogs (see Section \ref{sec:protostellar-reclassification} for details).}

\end{deluxetable*}

For line emission imaging, the continuum was first subtracted, along with only one change with respect to the above parameters, a modified version of the Briggs weighting scheme called \texttt{briggsbwtaper}, with the same robust value of $R=0.5$. This choice of weighting scheme modifies the cube imaging weights to have a similar density to that of the continuum imaging weights \citep{TheCasaTeam2022}, allowing for better results in our imaging. We use the $^{12}$CO line emission to search for evidence of outflows, which would then indicate the presence of a protostar.

All final images produced are corrected for primary beam attenuation: all figures presented in this paper are primary beam corrected. We visually identified any region of significant emission, and subsequently used \texttt{CASA}'s \texttt{imfit} task to fit 2D elliptical Gaussians to these regions.

\section{Detections}
\label{sec:detections}

\subsection{ALMA 12~m Detections}
\label{sec:alma-12m-detections}

We detect a total of 58 robust ALMA 12~m sources (hereafter 12~m detections); in addition, we find extended but significant areas of emission in the 12~m dataset: these eight ``extended" sources are marked \texttt{\_ext} in the 12~m data. Table \ref{tab:observed-properties-12m} lists each of the total 66 ALMA 12~m continuum sources with an index number, along with basic 2D Gaussian fitting properties for the robust detections. These include the peak and total integrated fluxes, along with the convolved and deconvolved source properties. We identify the sources that are marginally or fully unresolved with a value of -1 for their unresolved component(s) and the fitted size. For the extended sources, we only report both the peak flux and its associated location, as attempts at 2D Gaussian fitting were not viable.

\begin{deluxetable*}{lccDDDDDDrDDDDrr}
\digitalasset
\tablecaption{
Observed Properties of 12m Detections \label{tab:observed-properties-12m}
}

\tablehead{
    \colhead{Src} &
    \colhead{R.A.} &
    \colhead{Decl.} & 
    \multicolumn{2}{c}{Pk\tnm{a}} &
    \multicolumn{2}{c}{Pk$_{\text{err}}$\tnm{a}} &
    \multicolumn{2}{c}{Tot\tnm{a}} &
    \multicolumn{2}{c}{Tot$_{\text{err}}$\tnm{a}} &
    \multicolumn{2}{c}{FWHM$_{\text{a}}$\tnm{a}} &
    \multicolumn{2}{c}{FWHM$_{\text{b}}$\tnm{a}} &
    \colhead{P.A.\tnm{a}} &
    \multicolumn{4}{c}{FWHM$_{\text{a,d}}$\tnm{b}} &
    \multicolumn{4}{c}{FWHM$_{\text{b,d}}$\tnm{b}} &
    \multicolumn{2}{c}{P.A.$_{\text{d}}$\tnm{b}} \\
    \colhead{No.} &
    \colhead{(J2000)} &
    \colhead{(J2000)} & 
    \multicolumn{4}{c}{($\text{mJy}~\text{beam}^{-1}$)} &
    \multicolumn{4}{c}{(mJy)} &
    \multicolumn{4}{c}{(arcsec)} &
    \colhead{(deg)} &
    \multicolumn{4}{c}{(arcsec)} &
    \multicolumn{4}{c}{(arcsec)} &
    \multicolumn{2}{c}{(deg)} \\[-0.2cm]
    \colhead{} &
    \colhead{} &
    \colhead{} & 
    \multicolumn{4}{c}{} &
    \multicolumn{4}{c}{} &
    \multicolumn{4}{c}{} &
    \colhead{} &
    \multicolumn{2}{r}{fit} &
    \multicolumn{2}{r}{err} &
    \multicolumn{2}{r}{fit} &
    \multicolumn{2}{r}{err} &
    \colhead{fit} & 
    \colhead{err}
}

\decimals
\startdata
A1 & 18h25m13.32s & -02d59m55.00s & 2.51 & 0.22 & 2.76 & 0.41 & 1.89 & 1.35 & 92 & -1.00 & 1.26 & -1.00 & 0.55 & -1 & -1 \\
A2 & 18h27m54.23s & -03d42m41.50s & 4.67 & 0.25 & 11.04 & 0.82 & 2.44 & 2.20 & 154 & 1.99 & 0.23 & 1.49 & 0.24 & 168 & 23 \\
A3 & 18h27m54.73s & -03d42m38.40s & 4.76 & 0.24 & 5.12 & 0.43 & 1.68 & 1.46 & 112 & -1.00 & 0.69 & -1.00 & 0.58 & -1 & -1 \\
A4 & 18h28m09.08s & -03d48m10.88s & 38.09 & 0.52 & 72.29 & 1.41 & 2.11 & 1.98 & 152 & 1.60 & 0.06 & 1.14 & 0.07 & 175 & 9 \\
A5 & 18h29m02.58s & -01d38m57.97s & 9.85 & 0.24 & 19.27 & 0.68 & 2.41 & 1.86 & 90 & 1.73 & 0.10 & 1.25 & 0.08 & 84 & 9 \\
A6 & 18h29m03.63s & -01d39m06.06s & 1.86 & 0.21 & 10.73 & 1.39 & 4.23 & 3.11 & 100 & 3.88 & 0.54 & 2.80 & 0.39 & 100 & 24 \\
A7 & 18h29m05.33s & -01d41m56.91s & 17.89 & 0.27 & 19.48 & 0.50 & 1.75 & 1.42 & 104 & 0.54 & 0.13 & 0.36 & 0.21 & 127 & 51 \\
A8 & 18h29m05.46s & -03d42m45.90s & 1.40 & 0.20 & 2.01 & 0.46 & 2.41 & 1.38 & 96 & 1.69 & 0.71 & 0.32 & 0.43 & 94 & 29 \\
A9 & 18h29m07.14s & -03d43m24.13s & 2.64 & 0.21 & 7.34 & 0.76 & 2.93 & 2.18 & 54 & 2.52 & 0.34 & 1.51 & 0.32 & 46 & 15 \\
A10 & 18h29m08.18s & -01d30m47.25s & 4.29 & 0.25 & 7.49 & 0.64 & 2.32 & 1.74 & 96 & 1.58 & 0.23 & 1.09 & 0.19 & 94 & 28 \\
... & ... & ... & ... & ... & ... & ... & ... & ... & ... & ... & ... & ... & ... & ... & ... \\
\enddata

\tablecomments{The full version of this table is available in machine-readable format in the online journal. Only a portion is shown here for guidance regarding its form and function. The extended sources are shown below the robust sources with only their appropriate positions and associated peak flux values.}

\tablenotetext{a}{Properties of the Gaussian fit to the ALMA 12m emission: peak flux, integrated flux, major and minor axes of the FWHM, and position angle of the FWHM.}

\tablenotetext{b}{Properties of the deconvolved Gaussian fit: major and minor axes of the FWHM, and position angle of the FWHM. Fully unresolved sources are indicated by values of -1 in both the fit and err columns. For detected point sources and where possible, we give the upper limit for the deconvolved size in the err column, while leaving the fit column value of -1.}

\end{deluxetable*}

\subsection{ACA 7~m Detections}
\label{sec:aca-7m-detections}

We report a robust source catalog for the ACA 7~m dataset, based on a minimum detection level of 5$\sigma$ with respect to the reported rms values for each field in Table \ref{tab:observation-noise-levels}. Table \ref{tab:observed-properties-7m} shows the fitted properties for the 66 robust 7~m detections found in the dataset with this criteria.

\begin{deluxetable*}{lccDDDDDDrDDDDrr}
\digitalasset
\tablecaption{
Observed Properties of 7m ALAM-ACA Detections \label{tab:observed-properties-7m}
}

\tablehead{
    \colhead{Src} &
    \colhead{R.A.} &
    \colhead{Decl.} & 
    \multicolumn{2}{c}{Pk\tnm{a}} &
    \multicolumn{2}{c}{Pk$_{\text{err}}$\tnm{a}} &
    \multicolumn{2}{c}{Tot\tnm{a}} &
    \multicolumn{2}{c}{Tot$_{\text{err}}$\tnm{a}} &
    \multicolumn{2}{c}{FWHM$_{\text{a}}$\tnm{a}} &
    \multicolumn{2}{c}{FWHM$_{\text{b}}$\tnm{a}} &
    \colhead{P.A.\tnm{a}} &
    \multicolumn{4}{c}{FWHM$_{\text{a,d}}$\tnm{b}} &
    \multicolumn{4}{c}{FWHM$_{\text{b,d}}$\tnm{b}} &
    \multicolumn{2}{c}{P.A.$_{\text{d}}$\tnm{b}} \\
    \colhead{No.} &
    \colhead{(J2000)} &
    \colhead{(J2000)} & 
    \multicolumn{4}{c}{($\text{mJy}~\text{beam}^{-1}$)} &
    \multicolumn{4}{c}{(mJy)} &
    \multicolumn{4}{c}{(arcsec)} &
    \colhead{(deg)} &
    \multicolumn{4}{c}{(arcsec)} &
    \multicolumn{4}{c}{(arcsec)} &
    \multicolumn{2}{c}{(deg)} \\[-0.2cm]
    \colhead{} &
    \colhead{} &
    \colhead{} & 
    \multicolumn{4}{c}{} &
    \multicolumn{4}{c}{} &
    \multicolumn{4}{c}{} &
    \colhead{} &
    \multicolumn{2}{r}{fit} &
    \multicolumn{2}{r}{err} &
    \multicolumn{2}{r}{fit} &
    \multicolumn{2}{r}{err} &
    \colhead{fit} & 
    \colhead{err}
}

\decimals
\startdata
C1 & 18h25m13.07s & -02d59m52.56s & 4.35 & 2.11 & 3.07 & 3.16 & 13.80 & 7.75 & 97 & -1.00 & -1.00 & -1.00 & -1.00 & -1 & -1 \\
C2 & 18h27m54.30s & -03d42m40.79s & 24.51 & 2.88 & 34.19 & 6.32 & 16.55 & 12.61 & 83 & 8.96 & 4.37 & 5.86 & 3.52 & 48 & 59 \\
C3 & 18h28m09.07s & -03d48m10.77s & 93.80 & 3.99 & 103.22 & 7.46 & 14.68 & 10.91 & 92 & 4.43 & 2.04 & 2.08 & 2.72 & 24 & 86 \\
C4 & 18h28m44.67s & -01d30m36.66s & 7.32 & 2.30 & 5.08 & 3.32 & 13.03 & 8.20 & 86 & -1.00 & -1.00 & -1.00 & -1.00 & -1 & -1 \\
C5 & 18h28m45.11s & -01d26m54.68s & 10.89 & 2.69 & 10.80 & 4.77 & 15.16 & 9.83 & 110 & -1.00 & 13.26 & -1.00 & 5.81 & -1 & -1 \\
C6 & 18h29m02.53s & -01d38m57.42s & 22.55 & 3.07 & 22.22 & 5.41 & 14.83 & 9.93 & 95 & -1.00 & 9.79 & -1.00 & 3.19 & -1 & -1 \\
C7 & 18h29m03.51s & -01d39m05.82s & 12.38 & 3.23 & 18.83 & 7.68 & 19.84 & 11.45 & 90 & 13.45 & 8.75 & 5.03 & 4.15 & 86 & 21 \\
C8 & 18h29m05.30s & -01d41m55.42s & 22.72 & 3.72 & 33.54 & 8.49 & 17.14 & 12.87 & 123 & 11.19 & 5.27 & 3.92 & 6.59 & 150 & 44 \\
C9 & 18h29m07.09s & -03d43m23.96s & 10.85 & 2.21 & 13.12 & 4.42 & 15.94 & 11.00 & 101 & 6.91 & 7.54 & 3.97 & 3.61 & 124 & 52 \\
C10 & 18h29m08.16s & -01d30m47.13s & 12.81 & 3.61 & 16.96 & 7.63 & 15.84 & 12.54 & 153 & -1.00 & -1.00 & -1.00 & -1.00 & -1 & -1 \\
... & ... & ... & ... & ... & ... & ... & ... & ... & ... & ... & ... & ... & ... & ... & ... \\
\enddata

\tablecomments{The full version of this table is available in machine-readable format in the online journal. Only a portion is shown here for guidance regarding its form and function. The dim sources are shown below the robust sources with only their appropriate positions and associated peak flux values.}

\tablenotetext{a}{Properties of the Gaussian fit to the ALMA-ACA 7m emission: peak flux, integrated flux, major and minor axes of the FWHM, and position angle of the FWHM.}

\tablenotetext{b}{Properties of the deconvolved Gaussian fit: major and minor axes of the FWHM, and position angle of the FWHM. Fully unresolved sources are indicated by values of -1 in both the fit and err columns. For detected point sources and where possible, we give the upper limit for the deconvolved size in the err column, while leaving the fit column value of -1.}

\end{deluxetable*}

In addition to this formal catalog, we also report a secondary catalog with other structures that lie at lower significance levels, which we believe are real based on comparisons with the publicly available JCMT Gould Belt Survey 450~$\mu$m data taken with the SCUBA-2 instrument \citep[SCUBA-2 450~$\mu$m; ][]{Ward-Thompson2007, Kirk2018}. We guide our choices for the inclusion into this secondary catalog with the significance level of the emission itself and with visual checks of both the HGBS and SCUBA-2 450~$\mu$m maps for any correlation with morphology. In general, these cores lie at detection levels below 5$\sigma$, are dim, and are often quite extended in their morphology, making the fitting procedure inaccurate. There are a total of eight cores, which we mark as \texttt{\_dim}. The primary goal for this secondary catalog is to give as accurate fragmentation statistics in Section \ref{sec:multi-scale-analysis} as possible.

We continue the following sections with only the 12~m data being studied and bring in the 7~m data for the multiscale analysis in Section \ref{sec:multi-scale-analysis}.

\section{Protostellar Associations}
\label{sec:protostellar-associations}

Our analysis critically depends on accurate protostellar classification at two distinct spatial scales: (1) the dense cores revealed by Herschel ($\sim$\ang{;;18.2} or $\sim$7,900~au resolution), and (2) the compact substructures revealed  by our ALMA 12~m continuum detections ($\sim\ang{;;1}$ beam, $\sim$440~au resolution). These scales allow us to pose different questions: for the HGBS dense cores, we ask whether or not there are protostellar sources associated with these larger-scale structures identified by Herschel. The HGBS dense cores may have fragmented into multiple high-density peaks, which, if sufficiently bright and compact, may be detected with our ALMA 12~m observations. Thus, for the ALMA 12~m detections, we ask whether or not there are any known protostellar sources associated with each specific 12~m peak. In cases where there are multiple 12~m detections within a single HGBS core, it is therefore possible to have a protostellar HGBS core that contains both protostellar and starless 12~m detections. We will stick to using the terminology HGBS \textit{core} and 12~m \textit{detection} to try to make the distinction clear.

We associate the HGBS dense cores and our ALMA 12~m detections against three categories of complementary catalogs: infrared-based surveys (Section \ref{sec:ir-associations}), X-ray-based surveys (Section \ref{sec:xray-associations}), and millimeter-based multiwavelength surveys (Section \ref{sec:millimeter-associations}). We outline our methodology for cross-matching our ALMA 12~m detections in Section \ref{sec:alma-12m-associations-results}, and for cross-matching our HGBS dense cores in Section \ref{sec:protostellar-reclassification}, accounting for the inherent spatial resolution differences and appropriate association criteria at each scale.

\subsection{Associations with IR-based YSO catalogs}
\label{sec:ir-associations}

We searched through the literature and found six infrared catalogs that identified protostars in the Serpens/Aquila region, and we briefly describe each of these here.

The Wide-field Infrared Survey Explorer \citep[WISE;][]{Wright2010} provides all-sky mid-infrared photometry at 3.4, 4.6, 12, and 22 $\mu$m, with $\sim$\ang{;;6} resolution, offering broad coverage for identifying infrared excess sources. Specifically, we utilize the \citet{Marton2016} catalog, which combines WISE photometry with Two Micron All Sky Survey \citep[2MASS;][]{Skrutskie2006} and Planck dust opacity maps in order to classify YSOs.

For a more detailed look within the Gould Belt star-forming regions, we utilize the Spitzer cores-to-disks (c2d) and Gould Belt Legacy Survey compiled by \citet{Dunham2015}, which provide extinction-corrected spectral energy distributions and robust evolutionary classifications for YSOs based on combined IRAC ($3.6 - 8.0\mu$m) and MIPS ($24 - 160\mu$m) photometry at \ang{;;2} resolution.

Extending to the galactic midplane, the SPICY catalog \citep{Kuhn2021} employs machine learning techniques to identify YSO candidates from Spitzer/IRAC data, validated through variability analysis and spatial clustering. SPICY's coverage is limited to the galactic midplane ($|b| <1\degr$--$3\degr$), complementing rather than overlapping with the nearby cloud surveys of the extension of the Herschel Orion Protostar Survey (eHOPS) and c2d/GBS. While WISE provides all-sky coverage that encompasses all of these regions, its coarser angular resolution (\ang{;;6} compared to Spitzer/IRAC's $\sim$\ang{;;2}) limits its effectiveness in crowded regions and for detecting faint YSO candidates.

The Spitzer Extended Solar Neighborhood Archive\footnote{The data is publicly available at: \url{http://bit.ly/sesna2021}} \citep[SESNA; R. Gutermuth et al. 2026, in preparation;][]{Pokhrel2020,Pokhrel2023} is a uniform re-analysis of archival Spitzer surveys (resolutions of $\sim$\ang{;;2}), while additionally combining 2MASS observations to accurately classify the YSOs within dusty circumstellar material \citep{Pokhrel2020}. In addition, we use eHOPS \citep{Pokhrel2023}, which identifies a total of 172 protostellar sources in the Aquila region, tightly concentrated in the filamentary structures that permeate the overall cloud structure (also at a resolution of \ang{;;2}).

Finally, we also utilize the W40 multiwavelength study of \citet{Mallick2013}, which provides a comprehensive YSO catalog for the W40 region using UKIRT (United Kingdom Infrared Telescope) near-infrared $J$,$H$,$K$ bands (resolutions of $\sim$\ang{;;0.5}), Spitzer/IRAC, and Herschel PACS observations, complemented by $2.12\mu$m H$_2$ narrowband and GMRT radio continuum data.

\subsection{Associations with X-Ray-based YSO catalogs}
\label{sec:xray-associations}

The Massive Young Star-Forming Complex Study in Infrared and X-rays \citep[MystIX;][]{Feigelson2013, Povich2013} provides a YSO catalog constructed from multiwavelength data, combining infrared imaging and photometry from Spitzer/IRAC, 2MASS, and UKIRT with X-ray detections from the Chandra X-ray Observatory. Specifically, we use the candidate protostars catalog from \citet{Romine2016}.
This catalog identifies sources via two complementary pathways: infrared excess emission (by disk-bearing protostars) and X-ray detection (from evolved pre-main-sequence stars exhibiting coronal activity or strong stellar winds), with members requiring evidence from one of both. The angular resolution is primarily set by Chandra's subarcsecond X-ray imaging ($\sim$\ang{;;0.5}) and Spitzer/IRAC's infrared data ($\sim$\ang{;;2}).

\subsection{Associations with Additional Catalogs}
\label{sec:millimeter-associations}

\citet{Maury2011} carried out 1.2~mm dust continuum mapping of the Aquila rift complex with MAMBO (MAx-Planck Millimeter BOlometer array) on the IRAM 30~m telescope, resulting in the detection of 77 millimeter continuum sources. For each of these 77 millimeter continuum sources, \citet{Maury2011} folded in 2MASS $K$-band data, Spitzer 8 and 24~$\mu$m data, and Herschel $70-500$~$\mu$m to determine the spectral energy distribution. Sources which showed compact emission at or below 160~$\mu$m were classified as protostellar.\footnote{The astute reader will have noticed that the infrared observations used to determine the protostellar classifications here are the same as the data used in the catalogs described in Section \ref{sec:ir-associations}. Since the \citet{Maury2011} catalog uses infrared emission at the locations of millimeter cores, it is possible that faint new infrared sources could be identified using this method, adding protostars which were missed in the primary infrared catalogs described earlier. Since our goal is to ensure that we identify as many protostars as possible to have a clean sample of starless sources, we include the \citet{Maury2011} catalog where appropriate.}

The angular resolution with the IRAM 30~m antenna is $\sim\ang{;;11}$ at 1.2~mm. Consequently, the \citet{Maury2011} catalog is suitable only for associations with the HGBS dense cores, and has insufficient resolution to use for associations with the ALMA 12~m detections.

\subsection{ALMA 12~m Associations}
\label{sec:alma-12m-associations-results}

For classifying our ALMA 12~m detections, we only use primary infrared and X-ray protostellar catalogs with well-matched resolutions. As outlined in Section \ref{sec:millimeter-associations}, we specifically exclude the \citet{Maury2011} continuum millimeter catalog for associations with our 12~m detections as the spatial resolution is insufficient in the \citet{Maury2011} catalog.

Table \ref{tab:12m-detections-associations} summarizes the protostellar associations for each ALMA 12~m detection, alongside the results of our supplemental CO outflow search (see Section \ref{sec:alma-co-data}) and the final starless or protostellar classification. For each detection, we search for catalog sources within a radius cutoff, reporting the closest source from each catalog. The first two columns report the closest association from the SESNA and eHOPS catalogs, respectively, as these are the most complete catalogs for this region and are given the higher priority in our association methodology. The third column reports the closest association from the remaining catalogs in our search, namely the MystIX catalog \citep{Povich2013}, the W40 catalog \citep{Mallick2013}, and the SPICY catalog \citep{Kuhn2021}, and is only populated when no association is found in the first two columns. Together, these three columns capture all candidate protostellar associations within the search radius for a given detection.

\begin{deluxetable*}{lcDcDcDccDcc}
\digitalasset
\tablecaption{
12m Detections and Nearest Catalog Objects \label{tab:12m-detections-associations}
}

\tablehead{
    \colhead{Detection} &
    \multicolumn{3}{c}{SESNA\tnm{a}} &
    \multicolumn{3}{c}{eHOPS\tnm{b}} &
    \multicolumn{4}{c}{Other Catalogs\tnm{c}} &
    \multicolumn{3}{c}{Unique Association\tnm{d}} &
    \colhead{Outflow?\tnm{e}} & 
    \colhead{Final}\\
    \colhead{Name} &
    \colhead{Name} & \multicolumn{2}{r}{(arcsec)} & 
    \colhead{Name} & \multicolumn{2}{r}{(arcsec)} & 
    \colhead{Name} & \multicolumn{2}{r}{(arcsec)} & \colhead{References} &
    \colhead{Name} & \multicolumn{2}{r}{(arcsec)} &
    \colhead{} &
    \colhead{Classification}
}

\decimals
\startdata
A1 & J182513.33-025955.0 & 0.1 & aql-2 & 0.4 & ... & ... & ... & J182513.33-025955.0 & 0.1 & ... & protostellar \\
A2 & J182754.72-034238.5 & 8.0 & aql-6 & 0.9 & ... & ... & ... & aql-6 & 0.9 & ... & protostellar \\
A3 & J182754.72-034238.5 & 0.2 & aql-7 & 0.2 & ... & ... & ... & J182754.72-034238.5 & 0.2 & ... & protostellar \\
A4 & J182809.09-034812.5 & 1.7 & aql-9 & 1.5 & ... & ... & ... & aql-9 & 1.5 & ... & protostellar \\
A5 & J182901.84-013902.7 & 12.1 & aql-27 & 0.2 & ... & ... & ... & aql-27 & 0.2 & ... & protostellar \\
A6 & J182901.84-013902.7 & 27.1 & aql-29 & 1.0 & ... & ... & ... & aql-29 & 1.0 & ... & protostellar \\
A7 & J182905.32-014156.8 & 0.2 & aql-31 & 0.4 & ... & ... & ... & J182905.32-014156.8 & 0.2 & ... & protostellar \\
A8 & J182905.45-034245.6 & 0.3 & aql-32 & 0.6 & ... & ... & ... & J182905.45-034245.6 & 0.3 & ... & protostellar \\
A9 & J182905.45-034245.6 & 46.1 & aql-36 & 1.8 & ... & ... & ... & aql-36 & 1.8 & ... & protostellar \\
A10 & J182906.53-013103.0 & 29.3 & aql-38 & 0.9 & ... & ... & ... & aql-38 & 0.9 & ... & protostellar \\
... & ... & ... & ... & ... & ... & ... & ... & ... & ... & ... & ... \\
\enddata

\tablecomments{The full version of this table is available in machine-readable format in the online journal. Only a portion is shown here for guidance regarding its form and function.}

\tablenotetext{a}{Catalog entry name and separation distance to the nearest YSO entry in the SESNA catalog \citep[R. Gutermuth et al. 2026, in preperation;][]{Pokhrel2020}. We report the entry only if it lies within \ang{;;100} of the ALMA 12~m detection.}

\tablenotetext{b}{Catalog entry name and separation distance to the nearest YSO entry in the eHOPS catalog \citep{Pokhrel2023}. Again, we report the entry only if it lies within \ang{;;100} of the ALMA 12~m detection.}

\tablenotetext{c}{The ``Other Catalogs'' column only lists the catalog information for protostellar associations not covered by either the SESNA or eHOPS catalogs (see notes above). The References column indicates which catalogs have been used for these associations: (1) the MystIX catalog \citep{Povich2013}; (2) the multiwavelength W40 catalog \citep{Mallick2013}; (3) the SPICY catalog \citep{Kuhn2021}. Details can be found in Sections \ref{sec:ir-associations} and \ref{sec:xray-associations}.}

\tablenotetext{d}{This column reports the unique association that classifies each detection as protostellar (see Section \ref{sec:alma-12m-associations-results} for more details).}

\tablenotetext{e}{In the cases where the ALMA 12~m detection has no protostellar associations, we check the CO data for indications of an outflow. We report those three instances in this column (see Section \ref{sec:alma-co-data} for more details).}

\end{deluxetable*}

In some cases, the same protostar in one of the catalogs described above is the closest protostar to more than one ALMA 12~m detection. Since in reality, the protostar can only be associated with one unique ALMA 12~m detection, we include a fourth column in Table \ref{tab:12m-detections-associations} listing the closest unique protostellar association for each ALMA 12~m detection. In cases where multiple ALMA 12~m detections were initially matched to the same protostar, we take the closest 12~m detection as the one associated with the protostar. For the remaining 12~m detections, we search for other protostars located nearby that might be able to be associated with them. As a result, the fourth column in Table \ref{tab:12m-detections-associations} contains at most one protostar per ALMA 12~m detection, and each protostar appears at most once in the entire column. These uniquely associated protostars in the fourth column are used in our final classification of each ALMA 12~m detection as protostellar or starless.

When we examine the distribution of separations of the nearest protostar from each ALMA 12~m detection from the fourth column, we find a strong peak of separations of $<$~\ang{;;2} with a small tail of separations beyond \ang{;;10}. We interpret this distribution of separation as real protostellar associations within \ang{;;2} and random coincidence for the larger separations.

As indicated above, the majority of our 12~m detections are associated with protostars from the SESNA and eHOPS catalogs, accounting for a total of 45 protostellar associations. In addition, there are five other protostellar associations from other catalogs: sources A20 and A41 from the MystIX catalog \citep{Romine2016}, source A47 from the W40 catalog \citep{Mallick2013}, and sources A56 and A58 from the SPICY catalog \citep{Kuhn2021}.

In total, there are 50 ALMA 12~m detections directly associated with known protostellar sources, with the remaining 16 detections having no such infrared or X-ray associations. Table \ref{tab:12m-detections-associations} summarizes the protostellar associations found for each ALMA 12~m detection, as well as the results from a supplemental search for evidence of outflows from our CO data (see Section \ref{sec:alma-co-data}), and the final starless or protostellar classification.

\subsubsection{ALMA 12~m CO Data}
\label{sec:alma-co-data}

As noted in Section \ref{sec:alma-12m-band3-data}, one spectral window in our observations was configured for the line observation of $^{12}$CO$(1-0)$ at 115~GHz, hereafter CO. We utilize this data to search for hidden protostellar sources in our sample of 12~m detections without protostellar associations from our catalog searches. 
We find three such cases, all within the eastern area of observations near the galactic midplane, where only the SPICY catalog provides spatial coverage, limiting our ability to identify protostellar sources through protostellar catalog association.

Sources A54 and A55 are located in the HGBS dense core 183636.1-022144, with a separation between source A54 and A55 of approximately \ang{;;9.5} (or $\sim$4200~au). Both sources show prominent bipolar outflow signatures in CO, at almost perpendicular angles (see Figure \ref{fig:source55-source57-co} in Appendix \ref{sec:appendix-12co}).

Sources 57 and 58 form a comparable pair, located in the HGBS dense core 183640.3-023402 with a separation of approximately \ang{;;14.5} (or $\sim$6300~au). While Source 58 is classified as protostellar by the SPICY catalog, source 57 lacks a protostellar catalog association and would naively be considered starless; however, both exhibit bipolar outflow signatures in the CO data (see Figure \ref{fig:source55-source57-co} in Appendix \ref{sec:appendix-12co}).

Lastly, source A64\_ext shows marginal red-shifted CO emission in its vicinity; however, the emission morphology and spectral line profile are not clearly outflow-like and might instead be due to other structures in the envelope. While no protostellar catalog associations exist for this source, the HGBS data reveals a compact 70~$\mu$m source at the location of our detection. We nominally consider it a starless 12~m detection but note that it is a slightly ambiguous case.\footnote{For reference for the reader returning to this section after the later analyses, since source A64\_ext inhabits a core that is classified as protostellar by the HGBS catalog, it is \textit{not} counted as a starless 12~m detection within a starless dense core (Section \ref{sec:predicted-detection-count}).} We present the spectral details of A64\_ext (see Figure \ref{fig:source64_ext-co}) alongside all other sources discussed in this section in Appendix \ref{sec:appendix-12co}.

\subsubsection{ALMA 12~m Associations: Summary}
\label{sec:protostellar-summary}

Through our multiwavelength catalog searches, we directly associate 50 12~m detections to known protostellar sources, and in addition, we see evidence of outflowing material in CO from another three 12~m detections.

In total, our ALMA 12~m detections yield a total of 53 protostars, with the remaining 13 ALMA 12~m detections classified as starless. We discuss the starless detections in more detail in Section \ref{sec:candidate-starless-core-detections}.

\subsection{HGBS Core Reclassification}
\label{sec:protostellar-reclassification}

As outlined in Section \ref{sec:protostellar-associations}, our analysis depends on accurate classifications of the HGBS dense cores in addition to the ALMA 12~m detections.
Therefore, we perform similar check on the HGBS dense cores for any associations from more recent protostellar catalogs.

For these associations, we apply each HGBS core's deconvolved radius as the association criteria, similar to the half-power contour methodology originally implemented by \citet{Konyves2015}. To prevent spurious associations in extended cores, we set a maximum association radius of \ang{;;10}, corresponding to a few Jeans lengths at the typical HGBS core densities of $10^4-10^5$~cm$^{-3}$. For the associations, we use all of the protostellar catalogs discussed in Section \ref{sec:ir-associations}, \ref{sec:xray-associations}, and \ref{sec:millimeter-associations}.

Of the total 100 unstable cores observed with ALMA in our dataset, \citet{Konyves2015} classified 31 as protostellar based on the presence of pointlike 70~$\mu$m emission, which has been shown to trace well the internal luminosity of protostars \citep{Dunham2008}. Our new protostar catalog associations identify an additional 16 protostellar HGBS cores, bringing the total protostellar HGBS cores to 47, with the remaining 53 HGBS cores retaining their starless classification.

In detail, we reclassify as protostellar six prestellar cores given their proximity to SESNA sources, five prestellar cores given their proximity to \citet{Maury2011}-identified YSOs, three prestellar cores given their proximity to MystIX-identified YSOs \citep{Romine2016}, and two prestellar cores given their proximity to \citet{Mallick2013}-identified YSOs.

Table \ref{tab:observation-noise-levels} shows which observations have had their protostellar flag updated.

\subsection{Candidate Starless Substructure Detections}
\label{sec:candidate-starless-core-detections}

We report a total of 13 ALMA 12~m detections without direct protostellar associations or CO outflow signatures. Nine of these detections reside within protostellar HGBS cores, which have fragmented into a mixture of starless and protostellar substructures. We discuss the remaining four starless 12~m detections below.

\begin{figure*}
    \centering
    \includegraphics[]{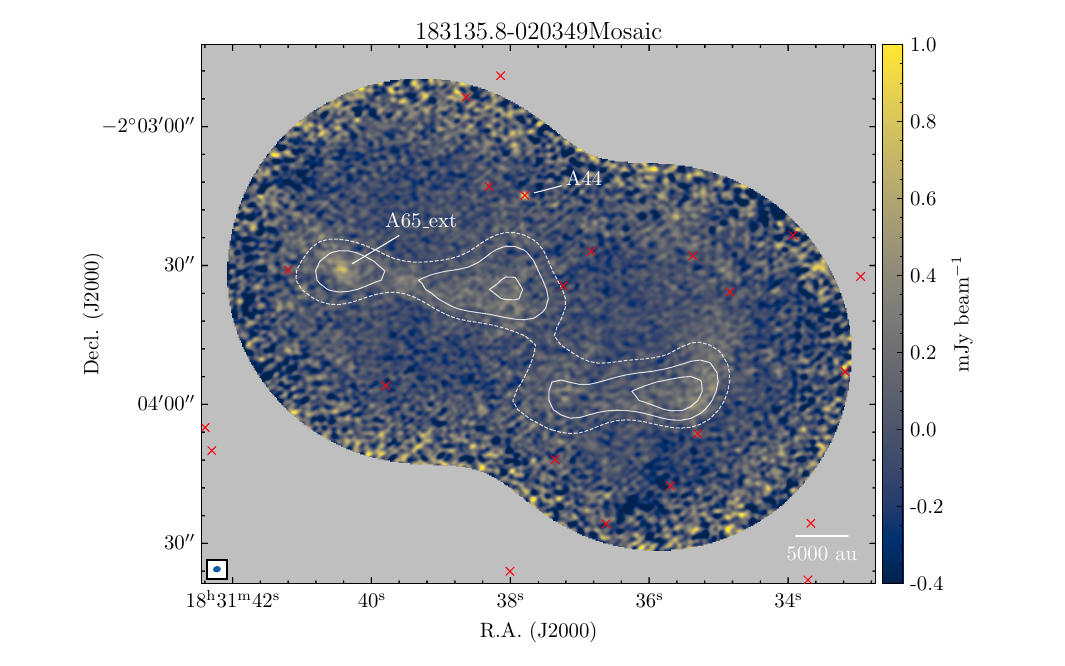}
    \caption{
        ALMA 12~m mosaic field 183135.8-020349Mosaic containing candidate starless core A65\_ext. The color bar ranges linearly from $-0.4-1.0$~mJy~beam$^{-1}$. The white contours correspond to the associated ACA 7~m continuum emission at the corresponding levels of $3\sigma$ (dashed lines), $5\sigma$, $7\sigma$, and $9\sigma$ (solid lines), where the $1\sigma$ rms is 1.0~mJy~beam$^{-1}$. All detections are labeled with their source number in white, and protostellar sources are plotted with red x markers. The synthesized beam is plotted in the bottom-left corner, along with a scale bar in the bottom-right corner indicating a linear distance of 5000~au, at the assumed distance to Aquila of 436~pc.
    \label{field-020349Mosaic-ref}
    }
\end{figure*}

Figure \ref{field-020349Mosaic-ref} shows extended source A65\_ext, a starless substructure candidate in the prestellar HGBS core 183139.2-020331 with a peak flux of 0.82~mJy~beam$^{-1}$. This source is found approximately $9\sigma$ above the noise floor, and is well characterized by coincident emission in the 7~m dataset.

\begin{figure*}
    \centering
    \includegraphics[]{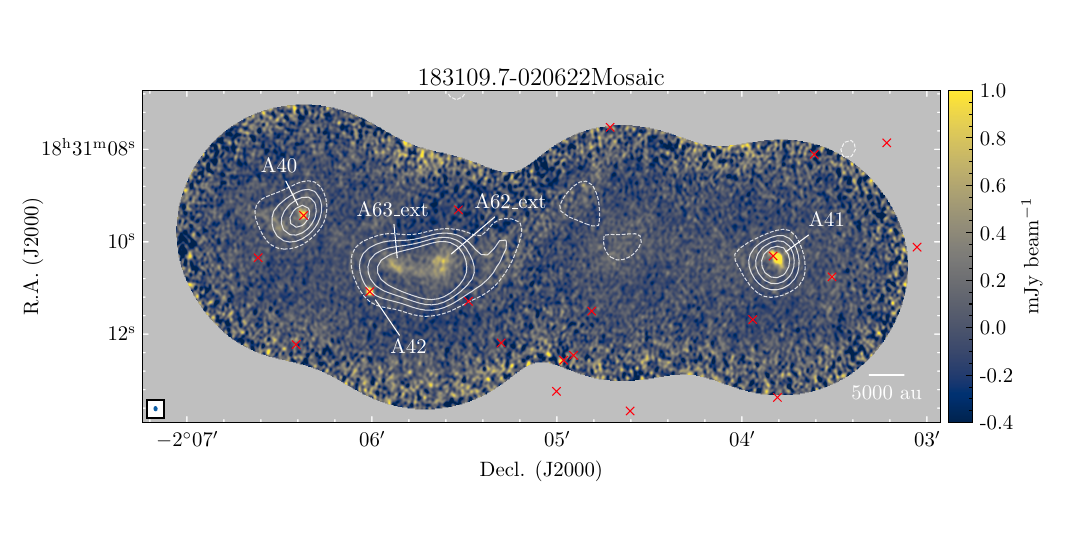}
    \caption{
    ALMA 12~m mosaic field 183109.7-020622Mosaic, with starless candidate sources A62\_ext and A63\_ext. See Figure \ref{field-020349Mosaic-ref} for plotting conventions. Note the unconventional axis orientation; the R.A. is displayed on the vertical axis.
    \label{fig:field-020622Mosaic}
    }
\end{figure*}

Figure \ref{fig:field-020622Mosaic} shows the extended sources A62\_ext and A63\_ext, both found within the prestellar HGBS core 183110.2-020542. The more northern substructure A62\_ext is marginally resolved into a few beam-sized structures\footnote{As shown in \citet{Caselli2019}, artificial subdivision into multiple substructures can be a natural consequence of interferometric observations of low-level emission. Due to the similarity between the beam size and the substructures detected, we caution against the view that this dense core structure is undergoing fragmentation; it may simply be a limitation due to the low amount of flux from the source.}, while the more southern source, A63\_ext, is an elongated structure in the southwest to northeast direction.

The final starless 12~m detection, source A64\_ext, presents a stranger case. As noted in Section \ref{sec:alma-co-data}, source A64\_ext is the only 12~m detection within an HGBS core classified as protostellar based on compact 70~$\mu$m emission. No additional protostellar catalog associations were found, and whilst CO emission was detected, it did not appear consistent with that expected from an outflow. Thus, we nominally have a protostellar HGBS core within which there is a single starless ALMA 12~m detection.

In summary, we detect three starless 12~m detections within two starless HGBS dense cores. This detection statistic is carried forward for our comparisons to simulations in Section \ref{sec:simulations-synthetic-observations}.

\section{Derived Properties of Detections}
\label{sec:derived_properties}

\subsection{Mass and Number Density Estimates}
\label{sec:mass-estimates}

We list the derived physical properties of each robust 12~m continuum source in Table \ref{tab:derived-properties-12m}, including the estimated mass, effective radius, and computed number density. Our effective radius is derived from the geometric mean of the semimajor and semiminor axes of the deconvolved size (see Table \ref{tab:observed-properties-12m}). Some sources are unresolved, and in these cases, the synthesized beam is used in place and is written as an upper limit.

An estimate of the mass of each continuum source is computed using the following equation:

\begin{equation}
    M = 100 \frac{d^2 S_{\nu}}{B_{\nu}(T_D)\kappa_{\nu}} \ ,
    \label{eq:mass-estimate}
\end{equation}

\noindent where $d$ is the distance to the cloud, $S_{\nu}$ is the integrated flux at the specific frequency $\nu$, and $B_{\nu}$ is the Planck function at the dust temperature of $T_D$. The factor of 100 represents the adopted gas-to-dust ratio. As discussed in Section \ref{sec:intro}, we adopt a distance of $d=436$ pc \citep{Ortiz-Leon2023}, along with a core temperature of $T=10$~K. While this temperature may be an underestimate for protostellar cores, we adopt a single temperature for simplicity in the mass calculations. We note that adopting a higher temperature would systematically reduce the derived masses and densities. For example, if a temperature of 20~K was adopted instead, the masses derived would be a factor of $\sim2.3$ lower.
We use the calculations of opacities from \citet{Ossenkopf1994} for an effective frequency of 106~GHz, which yields our choice in opacity of $\kappa_{\nu}=0.23$~cm$^2$~g$^{-1}$ (see \citetalias{Fielder2024} and references therein for a brief discussion on our choices).

We subsequently calculate the mean density of each ALMA 12~m source given the following:

\begin{equation}
    n = \frac{3}{4\pi \mu m_H}\frac{M}{R_{\text{eff}}^3} \ ,
    \label{eq:number-density}
\end{equation}

\noindent where $\mu = 2.80$ is the mean molecular weight per hydrogen molecule \citep{Kauffmann2008}. Table \ref{tab:derived-properties-12m} lists both the mass and number density estimates. In this case, the uncertainties are dominated primarily by systematic effects; these are much larger than the uncertainties in our 2D Gaussian fits.

For completeness, we also supply the same tabular information for our 7~m dataset; this can be found in Appendix \ref{sec:7m-derived-properties}.

\global\rotateonfalse
\global\deluxesidewaystablefalse
\global\deluxestarfalse
\begin{deluxetable*}{ccRcRc}
\digitalasset
\tablecaption{
Physical Properties of the ALMA 12m Detections \label{tab:derived-properties-12m}
}

\tablehead{
    \colhead{Src\tnm{a}} &
    \colhead{Mass\tnm{b}} & 
    \multicolumn{2}{c}{$R_{\text{eff}}$\tnm{b}} & 
    \multicolumn{2}{c}{Number Density\tnm{b}} \\ 
    \colhead{No.} &
    \colhead{($M_{\odot}$)} &
    \nocolhead{Lim.} &
    \colhead{(au)} &
    \nocolhead{Lim.} & 
    \colhead{($\text{cm}^{-3}$)}
}

\colnumbers
\startdata
A1 & 0.41 & < & 370 & > & 2.5e+08 \\
A2 & 1.65 & ... & 370 & ... & 9.5e+08 \\
A3 & 0.77 & < & 370 & > & 4.6e+08 \\
A4 & 10.80 & ... & 290 & ... & 1.3e+10 \\
A5 & 2.88 & ... & 320 & ... & 2.6e+09 \\
A6 & 1.60 & ... & 720 & ... & 1.3e+08 \\
A7 & 2.91 & ... & 100 & ... & 9.9e+10 \\
A8 & 0.30 & ... & 160 & ... & 2.2e+09 \\
A9 & 1.10 & ... & 420 & ... & 4.3e+08 \\
A10 & 1.12 & ... & 290 & ... & 1.5e+09 \\
... & ... & ... & ... & ... & ... \\
\enddata

\tablecomments{The full version of this table is available in machine-readable format in the online journal. Only a portion is shown here for guidance regarding its form and function. (3) and (5) show limit indicators for unresolved sources. For these sources, the synthesized beam has been used in-place for the deconvolved size; the effective radius should be taken as an upper limit, while the number density should be taken as a lower limit.}

\tablenotetext{a}{Running index number of the ALMA 12~m detections, the same as used in} Table \ref{tab:observed-properties-12m}.

\tablenotetext{b}{Estimated mass, effective radius, and number density. Unresolved sources are labeled with limit indicators, where the beam size has been used in-place of the size. Sources slightly larger than the beam size are deconvolved to sizes lower than the effective resolution; thus, caution should be used when interpreting those sizes.}

\end{deluxetable*}

\subsection{ALMA Peak and Integrated Flux}
\label{sec:alma-peak-flux}

\begin{figure}
    \centering
    \includegraphics[]{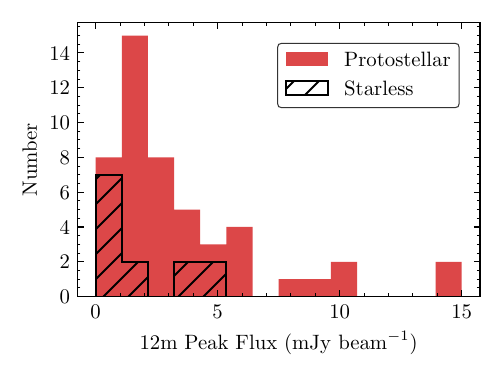}
    \caption{
    Distribution of ALMA 12~m detection peak fluxes. The population of 53 protostellar detections is plotted in red, while the 13 starless detections are plotted in black. Detections with a peak flux $>15$~mJy~beam$^{-1}$ are plotted in the final bin shown.
    \label{fig:alma-12m-peak-flux}
    }
\end{figure}

Figure \ref{fig:alma-12m-peak-flux} shows a histogram of the peak flux for the candidate starless and protostellar detections in our observations. In general, the distribution of protostellar sources extend to higher central densities, and, therefore, should have higher peak fluxes compared to their starless counterparts. The majority of our candidate starless core detections lie at lower peak flux bins agreeing well with both \citetalias{Fielder2024} and \citet{Kirk2017-Oph}.

\begin{figure}
    \centering
    \includegraphics[]{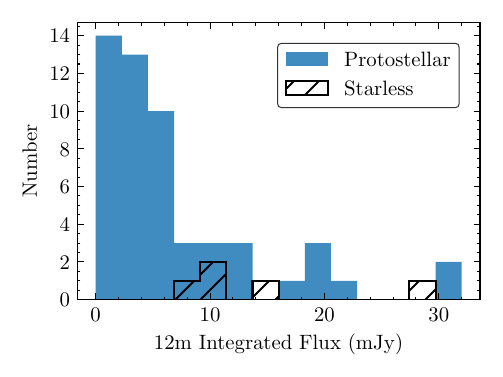}
    \caption{
    Distribution of ALMA 12~m detection total integrated fluxes. Detections with an integrated flux $>30$~mJy are plotted in the final bin shown. Due to the lack of Gaussian fits of the extended sources (\_ext), only five of the starless detections have reported data.
    \label{fig:alma-12m-total-flux}
    }
\end{figure}

Figure \ref{fig:alma-12m-total-flux} shows the distribution of integrated fluxes. Fitted parameters are available for only five starless detections, since the remaining sources are extended emission objects that could not be well characterized by our fitting procedure. Source A29 has one of the largest integrated fluxes of 29.06~mJy with the largest peak flux of the starless detections at 5.02~mJy~beam$^{-1}$.

\section{Numerical Simulations and Synthetic Observations}
\label{sec:simulations-synthetic-observations}

In order to interpret the number of starless core detections, comparisons with numerical simulations is helpful in discerning agreement with different core collapse models.
We use the same fundamental approach as \citet{Dunham2016}, \citet{Kirk2017-Oph}, and \citetalias{Fielder2024} to compute the expected number of starless core detections by the turbulent fragmentation model for the observed core sample in Aquila.

\subsection{Description of Simulations}
\label{sec:description-of-simulations}

We use magnetohydrodynamic simulations of isolated collapsing cores, to generate self-consistent, time-dependent physical conditions, to model the evolution of a dense core undergoing collapse. These starless cores are simulated using the ORION2 adaptive mesh refinement code base \citep{Li2021} and are the same as used in the previous studies \citep[\citetalias{Fielder2024}; ][]{Dunham2016, Kirk2017-Oph}. The simulations modeled two different core masses, 0.4 and 4.0~$M_{\odot}$; the previous studies used the 0.4~$M_{\odot}$ set of snapshots, as the typical masses of the observed dense core populations were closer to 0.4~$M_{\odot}$ than 4.0~$M_{\odot}$ \citepalias[e.g., the typical dense core mass in Orion B North's was 1.4~$M_{\odot}$ according to][]{Fielder2024}. The mean dense core mass in the Aquila HGBS catalog is 0.66~$M_{\odot}$, however, our observed subset of 100 dense cores is biased toward high-mass cores with a typical mass of 2.64~$M_{\odot}$. Therefore, we use the 4.0~$M_{\odot}$ set of snapshots for this study. We note that the results and conclusions are similar when using the 0.4~$M_{\odot}$ simulation due to similar detectability of the overdense regions within the simulated cores with the ALMA 12~m array \citepalias[see][for a similar discussion]{Fielder2024}.

We note that the $4.0~M_{\odot}$ simulation is identical to the M4P5 simulation of \citet{Offner2017} but with a higher number of AMR refinement levels permitted. While the M4P5 run of \citet{Offner2017} has a minimum resolution element of 26~au, the simulations presented here (and in the previous papers; see above) achieves a minimum resolution element of 3.3~au. We briefly summarize the simulation below and refer the reader to both \citet{Offner2017} and \citet{Dunham2016} for a more detailed description of the simulations.

The initial core contains $4.0~M_{\odot}$, has a radius of 0.065~pc ($n_{\rm H} \sim 10^5$~cm$^{-3}$), and has a temperature of 10~K. The core is threaded by a uniform supercritical magnetic field in the $z$-direction and is embedded in a warm low-density medium of 1000~K and density 100 times lower than that of the core. Initial turbulent velocity perturbations are applied to the gas, after which no further energy is injected. The collapse examined here occurs on scales well below those of filaments and large-scale feedback, and the simulations are evolved until a sink particle forms at densities of $\sim 4 \times 10^{11}$~cm$^{-3}$.

We generate flux maps of the simulations in the following way: (1) we re-grid the density cube to a common resolution of 20~au, (2) we project the cube into a total column density map along a direction perpendicular to the initial magnetic field ($z$-)direction, and (3) compute the flux map using Equation \ref{eq:mass-estimate} by utilizing the Planck function derived from the temperature in each pixel, along with the mass in each pixel.

We create synthetic observations of these flux maps using CASA's \texttt{simalma} task. We input the appropriate antenna configuration (C-3), total integration time (54.4~s), location in the sky (R.A.~=~18:30:30 and decl.~=~-02:00:00), precipitable water vapor (4.2~mm, which affects the noise calculations), and distance measurement for Aquila ($d=436$~pc) to mimic the real observations. In the same way, we utilize the same imaging parameters when invoking \texttt{tclean} for these synthetic measurement sets (see Section \ref{sec:calibration-reduction} for details).

\begin{figure*}
    \includegraphics[]{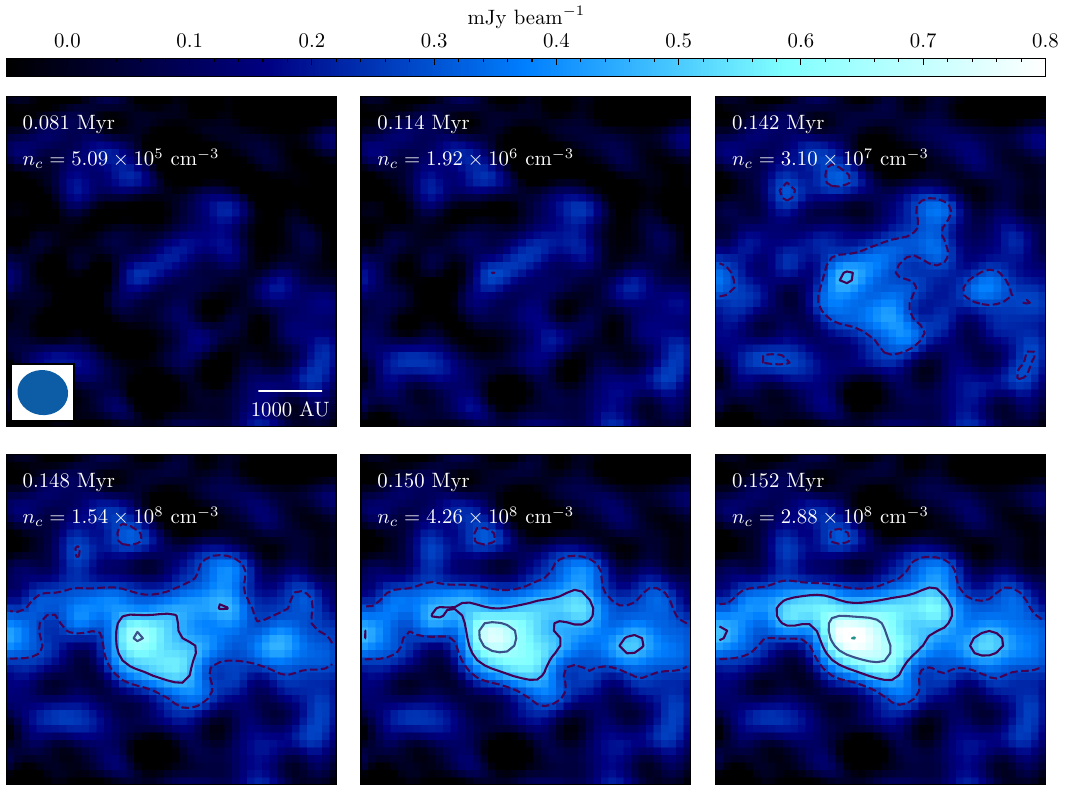}
    \caption{
    Synthetic ALMA 106~GHz observations of the 4.0~$M_{\odot}$ simulation, given at six time steps. Each panel indicates the simulation time, along with the central number density. The synthesized beam and scale bar are given in the first panel. Dashed contours represent the $3\sigma$ level of emission, while solid contours represent the $5\sigma$ level and increase by $2\sigma$, where $1\sigma$ rms $\sim$~0.08~mJy~beam$^{-1}$.
    \label{fig:40-simulations-436pc}
    }
\end{figure*}

Figure \ref{fig:40-simulations-436pc} shows a sample of six time steps, ranging from $0.081-0.152$~Myr of the synthetic 12~m images of the 4.0~$M_{\odot}$ simulation. These range from the earliest time step when the core begins to collapse, to the end of the simulation when a sink particle forms. We require a $5\sigma$ significance level for a positive detection, similar to the real observations, which occurs at 0.142~Myr when the core reaches a central number density of $3.10\times10^7$~cm$^{-3}$. For completeness, we note that we ran a similar test on the 0.4~$M_{\odot}$ simulation and found the $5\sigma$ detection threshold is \textit{very} similar, $2.48\times10^7$~cm$^{-3}$, which in turn yields near-identical predicted detections in the next section.

\subsection{Predicted Number of Starless Core Detections}
\label{sec:predicted-detection-count}

Similar to \citetalias{Fielder2024} and the previous studies, we compute the expected number of detections, $D$, with the assumptions of a continuous rate of star formation over a timescale as long as the typical core lifetimes:

\begin{equation}
    D > \frac{2}{3} \times N_{\text{total}} \times 
    \left( \frac{n_{\text{detectable}}}{n_{\text{limit}}}\right)^{-0.5} \ ,
    \label{eq:expected-detections}
\end{equation}

\noindent where $N_{\text{total}}$ is the number of prestellar cores observed, $n_{\text{detectable}}$ is the core density at which our ALMA observations can detect the core, and $n_{\text{limit}}$ is the observed lower limit of the mean core densities. 

The previous observations in the series \citep[\citetalias{Fielder2024}; ][]{Dunham2016,Kirk2017-Oph} were conducted on the full dense core population, providing an unbiased sample and single limiting minimum density. For this study, in order to increase the detection rate, gravitationally unstable cores were selected for observation (see Section \ref{sec:protostellar-reclassification}); however, this leads to a biased core sample without a single minimum density to use in Equation \ref{eq:expected-detections}.

\begin{figure*}
    \centering
    \includegraphics[scale=1.0]{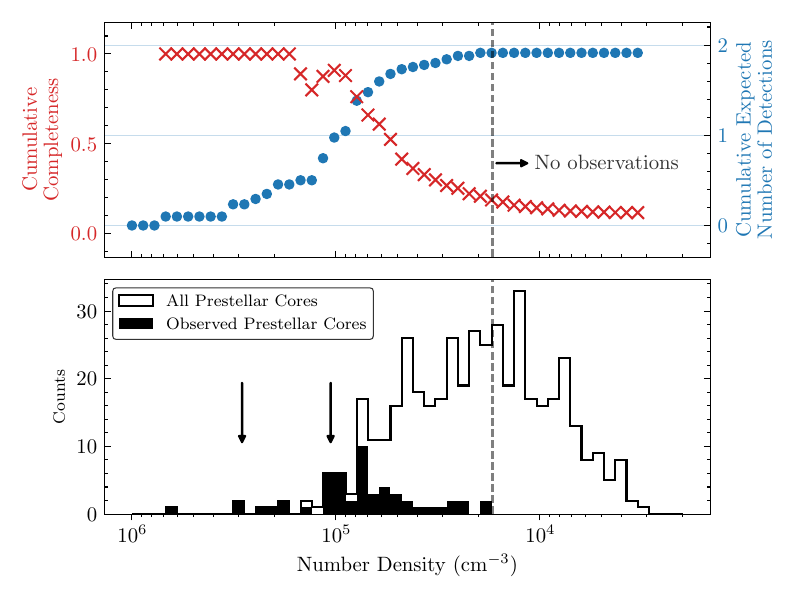}
    \caption{Top panel: cumulative completeness (red) and associated cumulative expected number of detections (blue; see Equation \ref{eq:expected_number_cumulative}) as a function of number density of the observed prestellar core population ($N=48$ cores). Bottom panel: the distribution of peak number density (peak flux divided by observed core area) of all prestellar cores in the HGBS dataset (line) and the ALMA observed prestellar cores (solid). A vertical dashed line in both plots marks the transition from the observed population to the unobserved population. 
    For context, vertical black arrows show the HGBS core densities that harbor our singly fragmented starless core detections.
    \label{fig:cumulative_expected_number_detections}
    }
\end{figure*}

Figure \ref{fig:cumulative_expected_number_detections} shows the peak number density distribution of the Aquila HGBS dataset, and the subset of ALMA observed cores. We therefore consider the effects of incompleteness in our sample at each chosen limiting observing density. We compute the expected number of detections across the density range of prestellar cores, according to the following:

\begin{equation}
    D_{\text{thresh}} > \frac{2}{3} \ \sum_{i=n_{\text{max}}}^{n_{\text{thresh}}} N_i \times
    \left( \frac{n_{\text{detectable}}}{n_i} \right)^{-0.5} \ ,
    \label{eq:expected_number_cumulative}
\end{equation}

\noindent where we cumulatively sum the expected number of detections per density bin $i$, starting from the highest-density bin $n_{\text{max}}$ and ending with our chosen threshold density bin, $n_{\text{thresh}}$.

Figure \ref{fig:cumulative_expected_number_detections} shows both the cumulative completeness percentage of our sample, along with the cumulative expected number of detections according to Equation \ref{eq:expected_number_cumulative}. At high number densities, there are fewer prestellar cores observed, and as a result, the expected number of detections is low. As we transition to lower number densities, our observed sample count increases, and the expected number of detections rise. Our observed dense core sample is 100\% complete above a density of $1.69\times10^5$~cm$^{-3}$, where we expect a total of 0.45 substructure detections. Around a number density of $7\times10^4$~cm$^{-3}$, the completeness sharply decreases, and as we no longer sample the majority of the cores, we are less confident about the expected number of detections expected in the observed sample. We adopt a threshold value of $7.91\times10^4$~cm$^{-3}$, in which we observe 32 of the total 42 cores at or above this number density, for a completeness of approximately 76\%.

To compute our expected number of detections, we adopt the above threshold value of $7.91\times10^4$~cm$^{-3}$, and a detectable density from our simulations of $3.10\times10^{7}$~cm$^{-3}$. Using Equation \ref{eq:expected_number_cumulative} along with standard Poisson statistics, we predict a minimum of 1 (1.38) $\pm$ 1 (1.1) prestellar HGBS cores that have the required $5\sigma$ detection significance above our chosen threshold density value, for a positive detection in our observations. 

As outlined in Section \ref{sec:candidate-starless-core-detections}, we identified three starless 12~m detections within starless HGBS cores. One of these lies within a starless HGBS core, while the remaining two share a single starless HGBS core. Therefore, we detect two starless HGBS cores that have starless 12~m detections within them. Since these two starless HGBS cores are found above the chosen threshold value (see Figure \ref{fig:cumulative_expected_number_detections}), we agree with the expected number of detections under the turbulent fragmentation collapse picture.

As an alternative to the turbulent core collapse mechanism, we contrast our results with the detectability of Bonnor-Ebert-like (BE) sphere models as approximations of traditional quasi-static models of core collapse. We use the \citet{Dapp2009} models to construct BE-sphere-like density profiles and compute the flux maps with appropriate temperature profiles (the same ones used for the construction of the turbulent simulation flux maps); more information can be found in Appendix \ref{sec:appendix-be-spheres}. We conclude that a thermally dominated dense core contracting to become a protostar with a BE-sphere-like density profile would not be detectable until the central density reaches approximately $10^{10}$~cm$^{-3}$, over 1000 times greater than turbulent simulation detectability threshold. Applying this detection threshold to Equation \ref{eq:expected_number_cumulative} yields a predicted number of detections of 0.08, lending confidence in the turbulent collapsing model as opposed to quasi-static core collapse models.

\subsection{Comparisons to Previous Studies}
\label{sec:comparison-to-previous-studies}

There are now four dedicated dense core surveys undertaken in a similar fashion with ALMA. The starless core observations in three distinct star forming regions: Ophiuchus, Orion B North, and Aquila, are all consistent with the turbulent fragmentation picture of core collapse \citep{Dunham2016, Kirk2017-Oph, Fielder2024}. \citet{Dunham2016} proposed that the lack of detections in Chameleon I (more consistent with a quasi-static core collapse model) was due to an inaccurate assumption of continuous star formation, and that the star formation rate in Chamaeleon I is declining, and cores are not currently collapsing. Supporting this idea, \citet{Tsitali2015} found that only five of the starless cores in Chamaeleon I (9\%) were gravitationally bound and that most structures in Chamaeleon may only be transient structures. Together, these results suggest that in actively star-forming regions, turbulence can help to shape the internal core structure, resulting in compact, higher-density zones at earlier times, which can be detectable with interferometers such as ALMA.

\section{Multiscale Analysis}
\label{sec:multi-scale-analysis}

To understand how mass is distributed on different spatial scales, we use several datasets to study the 53 prestellar and 47 protostellar cores observed in Aquila. We use the original HGBS observations \citep{Konyves2015}, our ALMA 12~m observations, and our 7~m observations for this analysis. We thus span spatial scales ranging from approximately $10^4$~au down to a few hundreds of astronomical unit, which, at the typical core masses in our datasets equate to number densities ranging from $10^5-10^{10}$~cm$^{-3}$. We include all 12~m \texttt{\_ext} sources and 7~m \texttt{\_dim} sources in our analysis, as we have visually inspected the SCUBA-2 450~$\mu$m data to confirm their existence.\footnote{Included in Appendix \ref{sec:appendix-supplementary-multiscale} is a figure set containing cutouts of the HGBS column density map, the ACA 7~m observation and the ALMA 12~m observations, for all 100 dense cores observed in Aquila. We also include a panel showing the SCUBA-2 450~$\mu$m emission; it has similar but slightly better angular resolution (\ang{;;9.6}) than the ACA 7~m dataset, and has generally higher signal to noise due to the substantially longer integration times. Since the JCMT 450~$\mu$m observations are at a wavelength more than 10 times shorter than our ALMA observations, the structures traced by the emission from the two facilities may not be identical.}

The basic theory of collapsing dense cores shows that the density of objects increases during the formation of the centrally embedded object; however, processes as simple as thermal feedback and as complex as turbulence can affect the fragmentation of the gas into a number of distinct objects. Here, we follow all of our cores over several distinct spatial scales to understand the nature of fragmentation as cores undergo collapse.

\subsection{Fragmentation Groups}
\label{sec:frag-groups}
We start by splitting our 100 observations of HGBS dense cores into five distinct fragmentation groups:

\begin{itemize}
    \item \textbf{Group A:}
    Dense cores that have neither detections in 12~m nor 7~m;
    \item \textbf{Group B:}
    Dense cores that have \textit{single} detections in 7~m but no detections in 12~m;
    \item \textbf{Group C:}
    Dense cores that have \textit{single} detections with 12~m but no detections in 7~m (the inverse of Group B);
    \item \textbf{Group D:}
    Dense cores that have \textit{single} detections in both 12~m and 7~m. These are known as the ``one-to-one'' group;
    \item \textbf{Group E:}
    Dense cores that have multiple detections in at least one of the 7~m or 12~m data.
\end{itemize}

We interpret Group A cores as those that have not yet undergone significant gravitational collapse (and may or may not in the future), while Group B has undergone only some gravitational collapse, enough to be detectable with the compact array only. We interpret Group D cores as simple monolithically collapsing structures, while Group E cores are fragmenting as they collapse. Group C cores are more puzzling on first glance, as they are only detected with the main array and not with the compact array; these will be discussed in Appendix \ref{sec:appendix-alma-12m-only-detections}. While young protostars are usually easily detected in our ALMA 12~m data, a small number of Group A and Group B cores are classified as protostellar. We examine these cores in detail in Appendix \ref{sec:appendix-protostellar-only-7m}.

Table~\ref{tab:group-based-statistics} summarizes our group-based population statistics. The largest group is Group A, the nondetections, which account for almost half of the observed population. The second largest group is Group D, accounting for just over a fifth of the dataset. Similarly, Group E, comprising complex cases, also accounts for around a fifth of the dataset.

\begin{deluxetable*}{ccchcc}
\tabletypesize{\small}
\tablecaption{
Multiscale Analysis: Group-based Statistics \label{tab:group-based-statistics}
}

\tablehead{
    \colhead{Group} &
    \colhead{Description} &
    \colhead{Total Number} &
    \nocolhead{Percentage of} &
    \colhead{Total Number of]} &
    \colhead{Total Number of} \\[-0.3cm]
    \nocolhead{} &
    \nocolhead{} &
    \colhead{of Cores} &
    \nocolhead{Total Cores} &
    \colhead{Prestellar Cores} &
    \colhead{Protostellar Cores} \\[-0.1cm]
    \nocolhead{} &
    \nocolhead{} &
    \colhead{($x$ / 100)} &
    \nocolhead{} &
    \colhead{($x$ / 53)} &
    \colhead{($x$ / 47)}
}

\startdata
    A & nondetected cores & 46 & 46\% & 39 & 7  \\
    B & 7m only detections   & 14 & 14\% & 10 & 4 \\
    C & 12m only detections  & 2  & 2\%  & 0 & 2 \\
    D & ``one-to-one'' detections & 21 & 21\% & 1 & 20 \\
    E & complex cases & 17 & 17\% & 3 & 14 \\
    \hline \\
    \multicolumn{6}{l}{Group E Breakdown} \\
     & & ($x$ / 17) & & ($x$ / 3) & ($x$ / 14) \\
    \hline
    E1 & single-level fragmenting cores & 8 & 8\% &  1 & 7 \\
    E2 & two-level fragmenting cores & 8 & 8\% &  2 & 6 \\
    E3 & others & 1 & 1\% &  0 & 1 \\
\enddata

\end{deluxetable*}

Notably, Group D contains a predominantly protostellar population of HGBS dense cores. In the previous surveys of Chamaeleon I \citep{Dunham2016}, Ophiuchus \citep{Kirk2017-Oph}, and Orion B North \citepalias{Fielder2024}, we detect only a few truly starless cores in the 12~m data. The prestellar cores that we do detect here mainly belong to either Group B (detected only in the 7~m data) or Group E (the complex cases).

We also present the statistics on two subsets of the fragmenting Group E cores, single-level fragmentation, or those that only have a single dense core at the 7~m scale (Group E1) or two-level fragmentation or those with multiple 7~m structures (Group E2). There is one leftover case, Group E3, in which a dense core fragments into two distinct 12~m cores without any intermediate 7~m detection; this case will also be discussed in Appendix \ref{sec:appendix-alma-12m-only-detections}.
While Group E2 has a higher incidence of starless/prestellar cores (2/8) than Group E1 (1/8), there is no statistically significant difference between the two.

\subsection{Fragmentation Hierarchy Overview}
\label{sec:heirarchies-overview}

\begin{figure*}
    \centering
    \includegraphics[]{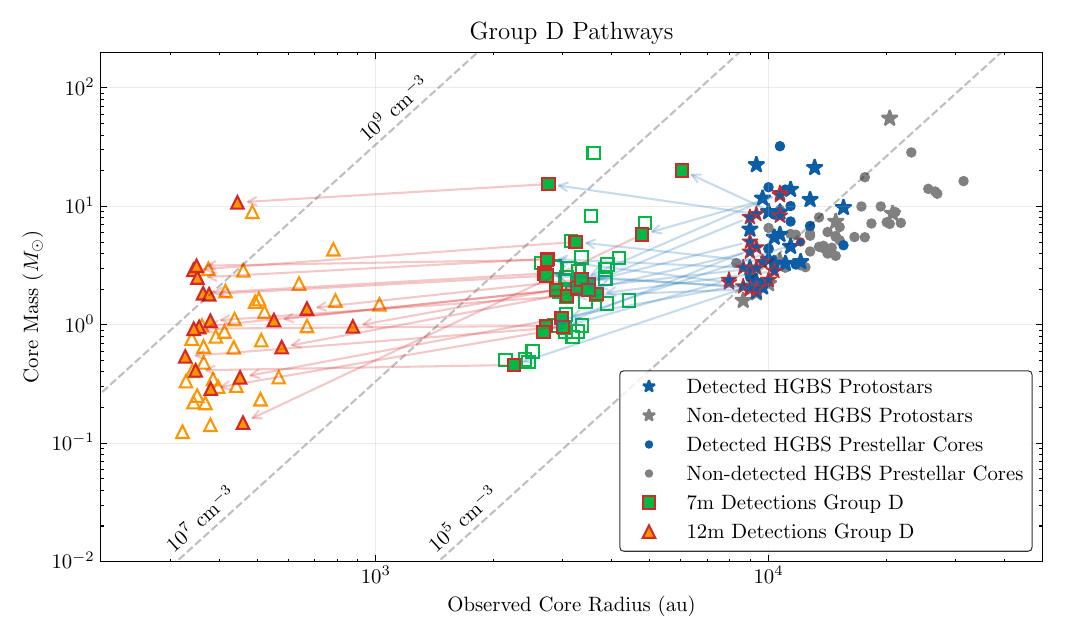}
    \caption{
    Mass vs. observed radius for the Group D cores. Blue arrows show the pathways between the HGBS cores and their associated 7~m substructures, while the red arrows show the pathways between the 7~m substructures and the 12~m substructures. The Group D cores across the three spatial scales are highlighted in a red border, while the other empty markers show the rest of the detections. In the HGBS dataset, circular markers show the prestellar core population, while the star-shaped markers show the protostellar population.
    \label{fig:multiscale-groupd-transitions}
    }
\end{figure*}

\begin{figure*}
    \centering
    \includegraphics[]{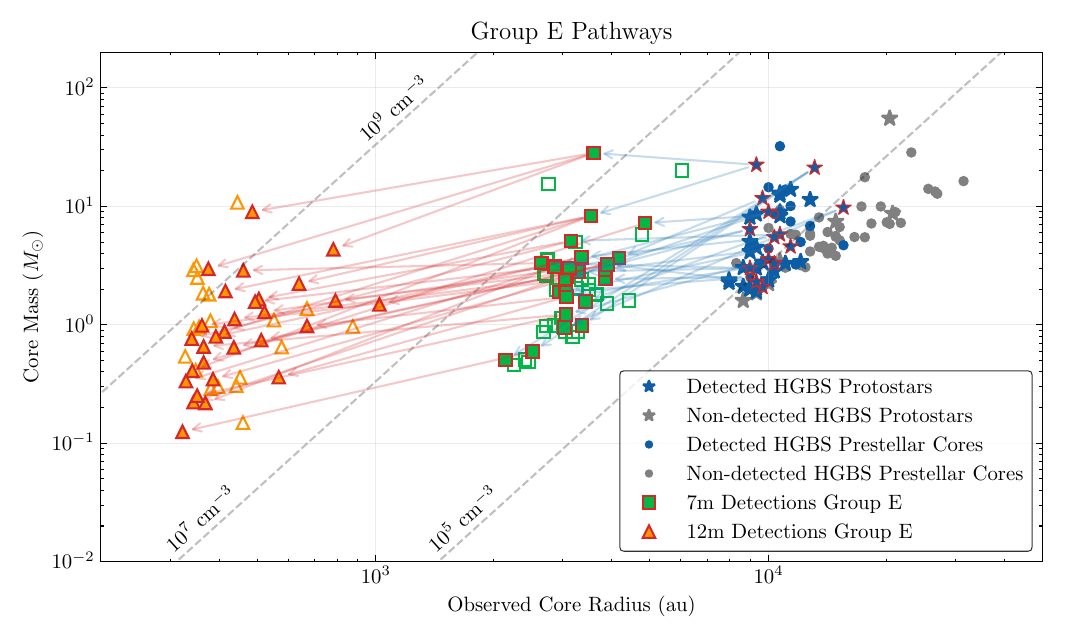}
    \caption{
    Mass vs. observed radius for the Group E cores. See Figure \ref{fig:multiscale-groupd-transitions} for plotting conventions.
    \label{fig:multiscale-groupe-transitions}
    }
\end{figure*}

Figures \ref{fig:multiscale-groupd-transitions} and \ref{fig:multiscale-groupe-transitions} show the underlying distribution of mass and observed core radius, of the original HGBS dense cores, along with the 7~m and 12~m detections. The HGBS dense cores have number densities of approximately $10^5$~cm$^{-3}$, with radii of tens of thousands of astronomical unit. The HGBS dense cores that are detected in the 7~m and 12~m data (plotted in blue) generally have smaller core radii. The 12~m data probes the smallest spatial scales of hundreds of astronomical unit, while the 7~m compact array data is sensitive to scales of thousands of astronomical unit.

As outlined in Section \ref{sec:detections}, only the peak fluxes are given for the extended detections in the 12~m and the dim detections in the 7~m. Mass estimates for these sources, therefore, cannot be computed, and this information is naturally left out of the following analysis.

\subsection{Fragmentation Pathways}
\label{fig:fragmentation-pathways}

In our observations of the 100 dense cores in Aquila, we detect a total of 54 of these cores with either or both the 7~m and 12~m data. We associate parent structures (those at larger spatial scales) to their child structures (those at smaller spatial scales), to see any trends in core populations across our different identified core groups (see Table \ref{tab:group-based-statistics}). We refer to the same cores (and their substructures) across different spatial scales as ``pathways".

Figure \ref{fig:multiscale-groupd-transitions} shows the pathways between different spatial scales for the Group D cores in our sample. The HGBS to 7~m pathways (as shown with the blue arrows) do not show any self-consistency. The HGBS to 7~m pathways are inherently complicated due to the nature of the different datasets. The 7~m and 12~m observations are at 3~mm, while the HGBS dense core catalog was created using a combination of data from $70-500$~$\mu$m. Comparing masses between the two sets of observations requires assumptions about the dust opacity and temperature. Additionally, differences in core boundary definitions will have a profound effect on the total integrated flux computed and, therefore the mass estimations. The pathways often show a decrease in mass from the HGBS spatial scale to the 7~m, but there are some instances where the mass increases, due to a combination of the reasons given above.

Conversely, the 7~m to 12~m pathways show clear self-consistency: a slight decrease in mass from the 7~m to the 12~m scale. Although the Group D 12~m substructures are biased toward smaller observed core radii, there is no underlying difference in the population compared to all other cores in all other groups. A two-sided Kolmogorov-Smirnov (K-S) test suggests that the properties of the Group D substructures at 12~m are randomly distributed among the full set of 12~m detections. On the other hand, the Group D 7~m substructures have a small bias toward smaller radii than the full set of Group D 7~m detections, with a two-sided K-S test showing a very weak difference in the populations ($p$=0.09).

Figure \ref{fig:multiscale-groupe-transitions} shows the same pathways between spatial scales, but for dense cores belonging to Group E. For the same reasons given above, there appears to be no consistent pathway from the largest HGBS spatial scale to the 7~m scale; however, a similar consistency is seen in the 7~m to 12~m pathway. Group E encompasses the complex cases, where multiple substructures are seen at the 7~m and 12~m spatial scales. As a result, while the pathways are steeper than the cores in Group D, there are often multiple substructures for each 7~m substructure. Additionally, the Group E substructure properties (mass, total flux, etc.) for both the 7~m and 12~m observations appear to be representative of (i.e., drawn randomly from) the full distribution of properties observed, according to K-S tests.

\begin{figure*}
    \centering
    \gridline{
        \fig{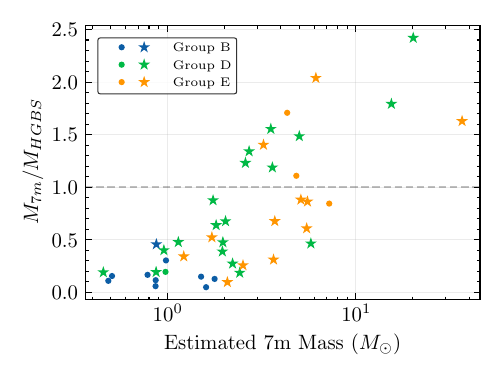}{0.5\textwidth}{}
        \fig{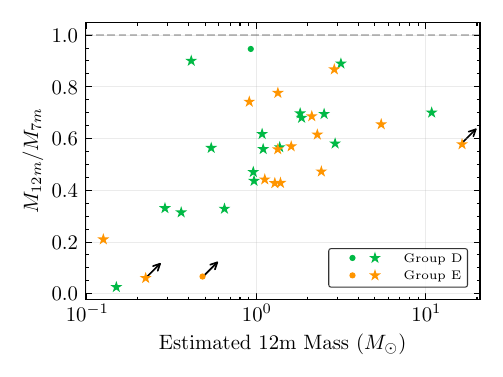}{0.5\textwidth}{}
    }
    \caption{
    The ratio of the substructure mass(es) to the parent mass, as a function of the total substructure mass. Circular markers show the prestellar core population, while the star-shaped markers show the protostellar population. Left panel: ACA 7~m detection(s) compared to their parent HGBS cores. Right panel: ALMA 12~m detections compared to their parent ACA 7~m cores. 
    For robust 12~m substructures that are accompanied by extended 12~m structure (of which we have no mass information), we label the nature of the lower limit with a diagonal gray arrow, showing the general movement in the plotting axes. The direction and length of the gray arrows have no significance.
    \label{fig:multiscale-missing-mass}
}
\end{figure*}

We compute the amount of mass estimated in the 7~m and 12~m observations for each core, summing together the multiple detections for sources in Group E. Figure \ref{fig:multiscale-missing-mass} shows the amount of estimated mass in all child structures as compared to the parent structures for both pathways studied. The left panel shows the HGBS to 7~m pathways for Groups B, D, and E. There are some instances where robust 12~m substructure are accompanied by extended 12~m substructure, in these cases, we indicate that the estimated masses reported are lower limits.

The majority of substructures do not recover the entire mass of the parent structure due to spatial filtering. In Figure \ref{fig:multiscale-missing-mass}, Group B cores stand out as having small estimated mass fractions: all have values of $<50$\%, with a median value of 17\%. This low estimated mass fraction is expected under our proposed scenario that these cores are at an early stage of gravitational collapse. At this stage, less mass is centrally concentrated, leading to a larger fraction of their flux being filtered out in the 7~m data and a complete lack of detection at 12~m. For Groups D and E, there appears to be a weak anticorrelation between 7~m mass/flux and the estimated mass fraction. This may be explained by extra filtering at the 7~m scale, as the most-massive 7~m cores in our sample are also the largest in spatial extent. The right panel of Figure \ref{fig:multiscale-missing-mass} shows the 7~m to 12~m pathways; all 12~m substructures recover less mass than the parent structure, as expected due to filtering affects. Most 12~m substructures only recover around $40$\%$-80$\% of the total mass found on the 7~m scale.

\subsection{Fragmentation and Multiplicity}
\label{sec:multiplicity}

Most stars form in multiple systems, and many of these systems form through gravitational fragmentation of the parent core \citep{Offner2023}. Instances of multiple protostars and/or substructures within a given core are a likely indication of a future multiple system. The observed multiplicity at different spatial scales may give insight as to whether multiplicity in substructure is inherited from larger spatial scales \citep{Pokhrel2018}. To compare multiplicity across all three spatial scales, we focus on the cores with complete data at each scale, specifically, those in Groups D and E. At our largest scales (i.e. the HGBS to 7~m pathway), the majority ($N=29$) of 7~m detections in our Group D and E cores are single-level fragmentation detections. There are eight total HGBS cores that are two-level fragmentation detections: seven HGBS cores with two 7~m substructures, and one core with three 7~m substructures. Therefore, the average number of substructures in our HGBS to 7~m pathway is 1.2.

As introduced in Section \ref{sec:intro}, \citet{Pokhrel2018} used five spatial scales across the Perseus molecular cloud to probe the number of fragments at each scale. Specifically, they use the Submillimeter Array and Very Large Array data in order to measure the observed multiplicity at the envelope scale as compared to the smaller disk and protostar scale. Similar to \citet{Pokhrel2018}'s analysis, we conduct a similar calculation for the 7~m to 12~m pathway, in which we compare the one-level fragmentation 7~m cores to the two-level fragmentation 7~m cores. We find the average amount of 12~m substructures to be 1.3 and 2.0, for the single-level and two-level fragmentation detections, respectively. \citet{Pokhrel2018} found a smaller but significant difference in the number of protostars in isolated envelopes (1.32) versus those found in grouped envelopes (1.47), consistent with our results in the one-level versus two-level fragmentation scenarios. Within their uncertainties, \citet{Pokhrel2018} stated that the population numbers in the isolated and grouped are equal, but that there may be a trend for grouped envelope structures to contain a larger amount of protostars. Despite our slightly smaller number populations, our data suggests that when conditions are such that fragmentation happens, it may happen at all scales. Numerical simulations further indicate that the extent and nature of fragmentation depend on local physical conditions, including magnetic field strength \citep[e.g.,][]{Lee2019, Guszejnov2023}, supporting the idea that the same environmental factors may regulate fragmentation from dense core to protostar scales.

\section{Conclusions}
\label{sec:conclusions}

We present 100 observations of prestellar and protostellar dense cores using ALMA's Band 3 correlator. We use these observations, and the resulting number of starless HGBS cores with starless 12~m detections, to test the turbulent core collapse scenario. We accomplish this using synthetic observations of turbulent magnetohydrodynamical simulations of starless core collapse, to probe at what densities substructures are detectable with ALMA's main (12~m) array. Additionally, we use three datasets at different representative spatial scales to study the subdivision of mass and its affect on the resulting fragmentation of dense core structures. We summarize our conclusions below.

We detect a total of 66 continuum sources with our 12~m data; three of these (starless) detections reside within two starless HGBS cores, which do not contain any protostellar substructures. We find an additional nine apparently starless detections that are adjacent to other nearby protostellar detections. In total, all 12 starless substructure detections lack any obvious outflow signatures in the $^{12}$CO line data, reinforcing their candidate starless nature. There is one other nominally starless detection that is a single fragment in a protostellar core, discussed in more detail in Appendix \ref{sec:appendix-12co}.

Using the same imaging parameters as our ALMA observations, we synthetically observe turbulent magnetohydrodynamical simulations of a collapsing starless core to predict how many detections we expect from such a model in our ALMA observations. We expect a minimum of $1 \pm 1$ detections while we observe two starless HGBS dense cores substructure. One of these HGBS dense cores harbors two candidate starless detections. Overall, we agree with the predictions from the turbulent core collapsing model, finding many candidate starless core substructures within starless dense cores. We also test the quasi-static core collapse model by observing critical BE-like sphere models. We show that the central density must reach $10^{10}$~cm$^{-3}$ before ALMA can robustly detect the dense core. In this scenario, we expect much less than one (0.07) detection in our ALMA observations, showing that our two detected dense cores with substructure and the respective turbulent model more accurately describes the core collapse scenario.

We include the ALMA-ACA 7~m dataset to study dense core substructures on three different spatial scales ranging from hundreds of astronomical unit to tens of thousands of astronomical unit. We find 17 dense cores that have highly fragmenting substructures within, some of which have a mix of starless and protostellar components. The dense cores with a mix of starless and protostellar components may be in a very interesting time of formation (see Appendix \ref{sec:appendix-adjacent-detections} for details). We find 14 dense cores that are not centrally concentrated enough to be observed with the 12~m data, and are only seen in the 7~m data. The aforementioned dense cores are more likely prestellar in nature, not having yet evolved to form protostars.

Despite choosing the most apparently gravitationally unstable systems for the observations presented in this study, nearly half of the dense cores have no detections with either the ALMA-ACA array or the ALMA main array, indicating that these have not yet begun collapsing or that they might not form stars. As suggested by turbulent simulations, most cores that reach central densities of $n_{\text{H}_2}\sim10^4$~cm$^{-2}$ do not go on to form protostars, while those that reach the threshold of the ALMA detections here are likely to do so \citep{Offner2022, Offner2025}.

Due to the spatial filtering of interferometers, we find that our 12~m observations only typically recover $40$\%$-80$\% of the total mass found in the parent 7~m substructures. Additionally, we find more fragmentation at the 12~m scale when there is more substructure at the 7~m scale, suggesting that when the conditions are right for fragmentation, it happens on all scales.

\begin{acknowledgments}
We thank the referee for providing insightful reports that greatly improved the manuscript.
S.D.F. acknowledges and respects the Lekwungen-speaking peoples on whose traditional territories S.D.F. works and lives, and the Songhees, Esquimalt, and WSANEC peoples whose historical relationships with the land continue to this day.
This paper makes use of ALMA data with project code Nos. 2018.1.00197.S and 2016.1.00320.S. ALMA is a partnership of ESO (representing its member states), NSF (USA), and NINS (Japan), together with NRC (Canada) and NSC and ASIAA (Taiwan), in cooperation with the Republic of Chile. The Joint ALMA Observatory is operated by ESO, AUI/NRAO, and NAOJ. 
This research also made use of NASAs Astrophysics Data System (ADS) Abstract Service. 
This research has made use of the SIMBAD database, operated at CDS, Strasbourg, France \citep{Wenger2000}. 
The JCMT has historically been operated by the Joint Astronomy Center on behalf of the Science and Technology Facilities Council of the United Kingdom, the National Research Council of Canada, and the Netherlands Organisation for Scientic Research. 
The authors acknowledge the use of the Canadian Advanced Network for Astronomy Research (CANFAR) Science Platform operated by the Canadian Astronomy Data Centre (CADC) and the Digital Research Alliance of Canada (DRAC), with support from the National Research Council of Canada (NRC), the Canadian Space Agency (CSA), CANARIE, and the Canadian Foundation for Innovation (CFI).
S.D.F. acknowledge the support of the Natural Sciences and Engineering Research Council of Canada (NSERC), 600061-2025. Cette recherche a été financée par le Conseil de recherches en sciences naturelles et en génie du Canada (CRSNG), 600061-2025. 
S.D.F. and H.K. also acknowledge the support from an NSERC Discovery Grant.
S.O. acknowledges support from NSF AAG-2407522, a Peter O’Donnell Distinguished Researcher Fellowship, and a Donald Harrington Fellowship.
\end{acknowledgments}

\vspace{5mm}
\facilities{ALMA, HST, JCMT (SCUBA-2)}
\software{\texttt{auto\_selfcal} \citep{PatrickSheehan2025}, Astropy \citep{AstropyCollaboration2013, AstropyCollaboration2018, AstropyCollaboration2022}, \texttt{CASA} \citep{TheCasaTeam2022}, SciencePlots \citep{JohnGarrett2023}}

\section*{Data Availability}
\label{sec:data-availability}

All data presented in this paper, including compiled figures and tables, are publicly available at the CADC via the following link: \dataset[DOI: 10.11570/25.0103]{https://doi.org/10.11570/25.0103} \citep{Fielder2026-Aquila-CADC-DOI}. In addition, the full suite of $^{12}$CO data cubes is also given for those interested.

\clearpage
\appendix

\section{ALMA-ACA 7~m Detections: Derived Properties}
\label{sec:7m-derived-properties}

For completeness, Table \ref{tab:derived-properties-7m} provides the derived properties of the 7~m data using the same assumptions as outlined in Section \ref{sec:mass-estimates}.

\begin{deluxetable}{ccRcRc}[h!]
\digitalasset
\tablecaption{
Physical Properties of the ALMA-ACA 7m Detections \label{tab:derived-properties-7m}
}

\tablehead{
    \colhead{Src\tnm{a}} &
    \colhead{Mass\tnm{b}} & 
    \multicolumn{2}{c}{$R_{\text{eff}}$\tnm{b}} & 
    \multicolumn{2}{c}{Number Density\tnm{b}} \\ 
    \colhead{No.} &
    \colhead{($M_{\odot}$)} &
    \nocolhead{Lim.} &
    \colhead{(au)} &
    \nocolhead{Lim.} & 
    \colhead{($\text{cm}^{-3}$)}
}

\colnumbers
\startdata
\global\colnumsused=0\relax 
C1 & 0.46 & < & 3230 & > & 4.1e+05 \\
C2 & 5.11 & ... & 1580 & ... & 3.9e+07 \\
C3 & 15.43 & ... & 660 & ... & 1.6e+09 \\
C4 & 0.76 & < & 3230 & > & 6.8e+05 \\
C5 & 1.61 & < & 3230 & > & 1.5e+06 \\
C6 & 3.32 & < & 3230 & > & 3.0e+06 \\
C7 & 2.81 & ... & 1790 & ... & 1.5e+07 \\
C8 & 5.01 & ... & 1440 & ... & 5.1e+07 \\
C9 & 1.96 & ... & 1140 & ... & 4.0e+07 \\
C10 & 2.53 & < & 3230 & > & 2.3e+06 \\
... & ... & ... & ... & ... & ... \\
\enddata

\tablecomments{The full version of this table is available in machine-readable format in the online journal. Only a portion is shown here for guidance regarding its form and function. (3) and (5) show limit indicators for unresolved sources. For these sources, the synthesized beam has been used in-place for the deconvolved size; the effective radius should be taken as an upper limit, while the number density should be taken as a lower limit.}

\tablenotetext{a}{Running index number of the ACA detections, the same as used in Table \ref{tab:observed-properties-7m} where we present the ACA 7~m detection observed properties.}

\tablenotetext{b}{Estimated mass, effective radius, and number density. Unresolved sources are labeled with limit indicators, where the beamsize has been used in-place for the size. Sources slightly larger than the beamsize are deconvolved to sizes lower than the effective resolution, thus caution should be used when interpreting those sizes.}

\end{deluxetable}

\section{Unique and Interesting Detections (and Nondetections) of Substructure}
\label{sec:appendix-unique-and-interesting}

\subsection{Adjacent Candidate Starless Core Substructure}
\label{sec:appendix-adjacent-detections}

In several systems, we identify adjacent candidate starless core substructures that may represent small clusters or groups in the process of formation. These groupings often show a combination of starless and protostellar components, suggesting that we are observing them during an early stage of formation. While coeval formation is often assumed for multiple systems, this may not always be true. For example, \citet{Murillo2016} examined protostellar systems in Perseus and found that roughly one-sixth to one-third of the observed SED classes could not be explained by coeval formation, even after accounting for orientation effects that can bias class determinations \citep[e.g.,][]{Robitaille2006, Robitaille2007, Dunham2008, Offner2012, Murillo2016}. This indicates that some systems may evolve asynchronously, though some apparent class differences may also reflect geometry rather than true age spreads.

The mixed populations we observe could therefore represent either non-coeval systems, where some fragments have already formed protostars while others remain starless, or systems caught just at the brink of formation, still evolving on similar physical timescales. Simulations by \citet{Generozov2025} suggest that the formation of binary and higher-order multiple systems occurs on timescales comparable to the freefall time of the gas and well within the protostellar lifetime, consistent with small temporal offsets within a single rapid formation event. In our observations, the protostellar members of these groups are predominantly Class 0/I, consistent with this interpretation. We highlight all of the systems identified in our observations here to facilitate future follow-up.

\begin{figure*}
    \centering
    \gridline{
        \fig{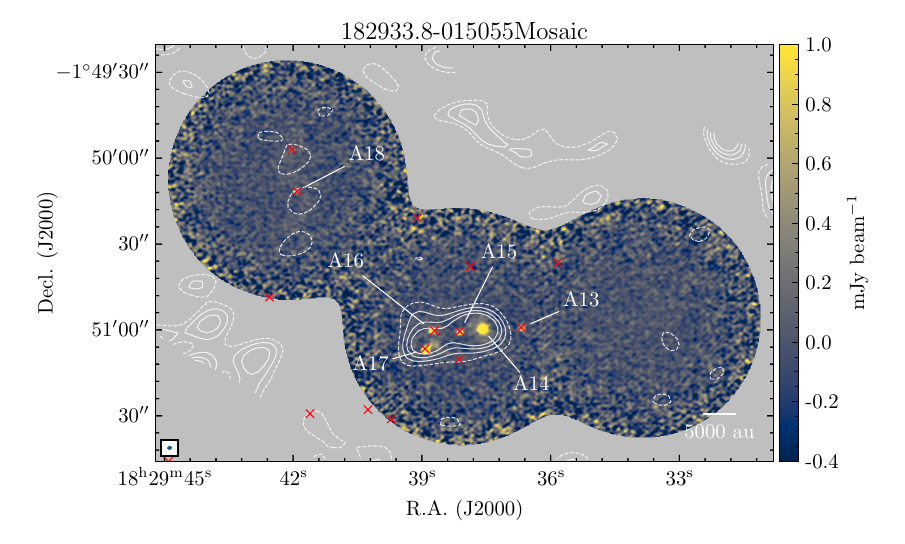}{0.85\textwidth}{}
    }
    \gridline{
        \fig{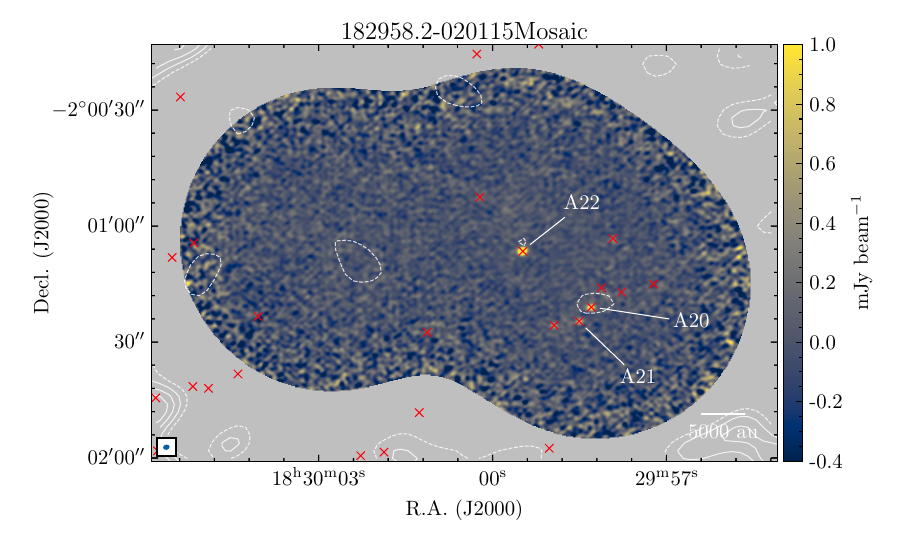}{0.85\textwidth}{}
    }
    \caption{
    This figure presents examples of adjacent candidate starless core substructures discussed in Section \ref{sec:appendix-adjacent-detections}, illustrating groupings of starless and protostellar components that may represent small clusters in the early stages of formation.
    Top panel: ALMA 12~m mosaic field 182933.8-015055Mosaic, with starless candidate source A14. Bottom panel: ALMA 12~m mosaic field 182958.2-020115Mosaic, containing the starless candidate source A20. See Figure \ref{field-020349Mosaic-ref} for plotting conventions.
    \label{fig:adjacent-starless-1}
    }
\end{figure*}

The top panel of Figure \ref{fig:adjacent-starless-1} shows source A14, a starless substructure component in the actively star-forming protostellar core 182938.0-015058, containing many other nearby detected protostars. Similarly, the bottom panel of Figure \ref{fig:adjacent-starless-1} shows source A20, a starless component of the HGBS 182958.2-020115 protostellar core also with many nearby protostars.

\begin{figure*}
    \centering
    \gridline{
        \fig{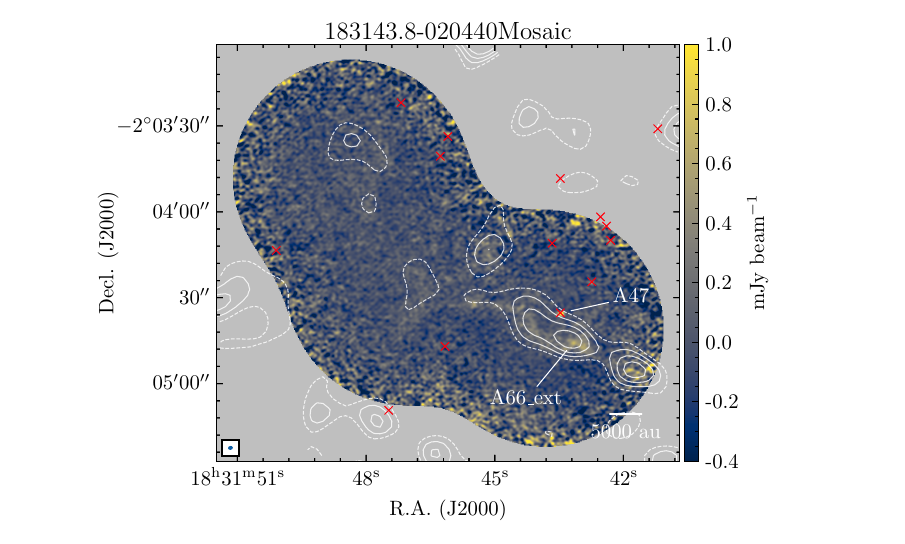}{0.85\textwidth}{}
    }
    \gridline{
        \fig{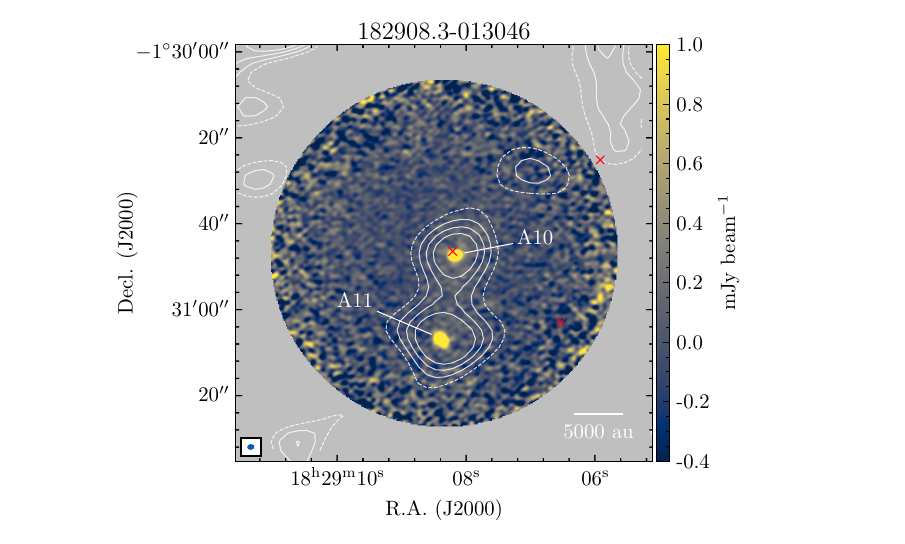}{0.85\textwidth}{}
    }
    \caption{
    This figure presents examples of adjacent candidate starless core substructures discussed in Appendix \ref{sec:appendix-adjacent-detections}, illustrating groupings of starless and protostellar components that may represent small clusters in the early stages of formation.
    Top panel: ALMA 12~m mosaic field 183143.8-020440Mosaic, with starless candidate source A66\_ext. Bottom panel: ALMA 12~m field 182908.3-013046, containing the starless candidate source A11. See Figure \ref{field-020349Mosaic-ref} for plotting conventions.
    \label{fig:adjacent-starless-2}
    }
\end{figure*}

The bottom panel of Figure \ref{fig:adjacent-starless-2} shows source A11, a starless component of the HGBS 182908.3-013046 protostellar core, while the top panel shows source A66\_ext, a nearby candidate starless component nearby protostellar source A47.

\subsection{ALMA 12~m Only Detections}
\label{sec:appendix-alma-12m-only-detections}

In three separate cases, we detect emission from HGBS dense cores from only the ALMA 12~m main array, with no comparable detection with the ALMA-ACA 7~m compact array. The three ALMA 12~m detections have very low peak fluxes, which fall well below the minimum $3\sigma$ detection threshold for the ACA. These substructures are thus sufficiently diluted in the larger 7~m beam and go undetected in our observations.

\begin{figure*}
    \centering
    \noindent
    \includegraphics[width=0.5\textwidth]{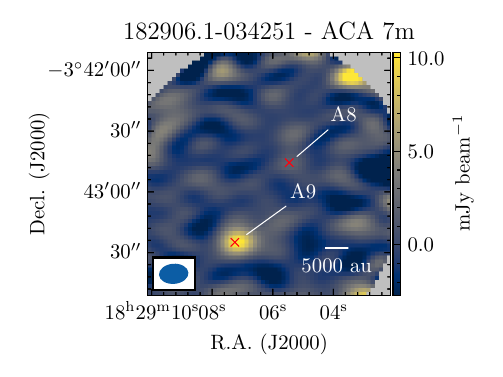}%
    \includegraphics[width=0.5\textwidth]{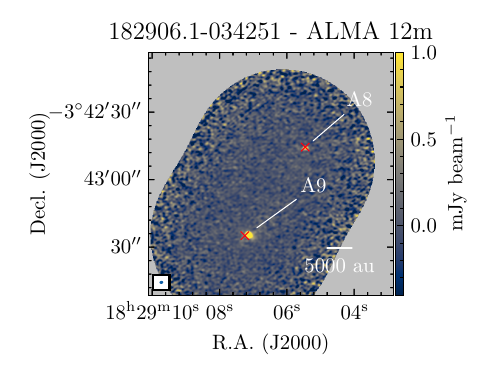}\\[0pt]
    \includegraphics[width=0.5\textwidth]{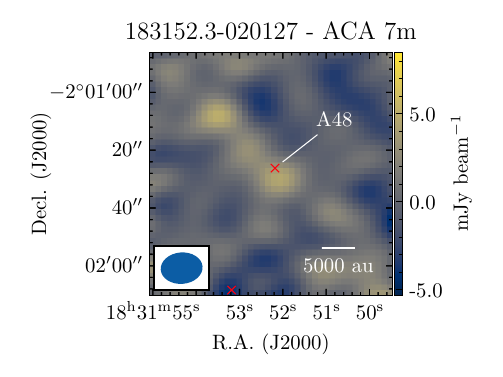}%
    \includegraphics[width=0.5\textwidth]{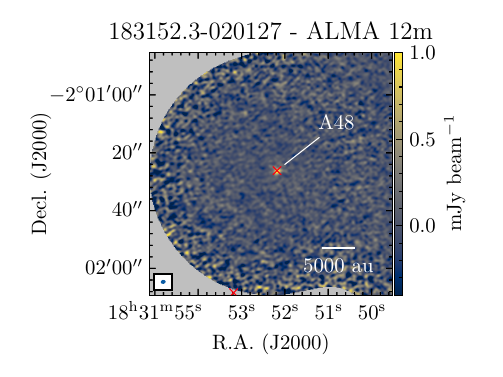}\\[0pt]
    \includegraphics[width=0.5\textwidth]{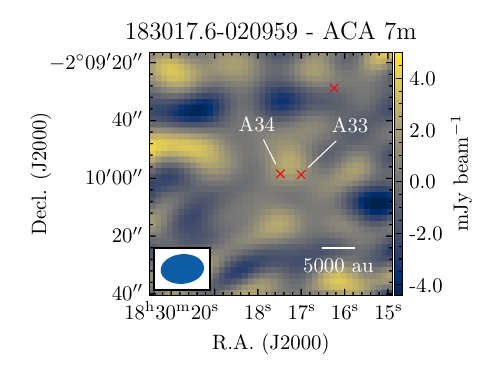}%
    \includegraphics[width=0.5\textwidth]{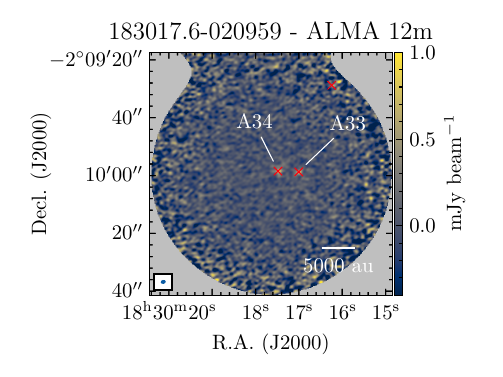}
    \caption{
    HGBS dense cores detection by the ALMA 12~m array (right panels), with no comparable detection with the ALMA-ACA 7~m array (left panels). Plotting conventions generally follow those found in Figure \ref{field-020349Mosaic-ref}.
    \label{fig:12m-only-detections}
    }
\end{figure*}

These three cases are presented in Figure \ref{fig:12m-only-detections}. The top row shows the HGBS dense core 182906.1-034251 which harbors the ALMA 12~m source A8, with a peak flux of 1.40~mJy~beam$^{-1}$. Our $3\sigma$ ACA sensitivity ($1\sigma$ rms = 0.84~mJy~beam$^{-1}$) is insufficient to detect this source. The middle row shows HGBS dense core 183152.3-020127 which contains ALMA 12~m source A48 with a peak flux of 1.20~mJy~beam$^{-1}$, whereas the ACA $1\sigma$ sensitivity is comparable with a value of 1.19~mJy~beam$^{-1}$. Finally, the bottom row shows HGBS dense core 183017.6-020959 containing the ALMA 12~m sources A33 and A34 with peak fluxes of 0.76 and 0.70~mJy~beam$^{-1}$, respectively. The associated ACA $1\sigma$ sensitivity for this observation is larger with a value of 1.00~mJy~beam$^{-1}$.

\subsection{Potential Streamer-like Morphologies}
\label{sec:appendix-streamer-like-morphologies}

Streamers, narrow velocity-coherent structures, have emerged as particularly intriguing features in star-forming regions, as they appear to play a role in channeling material onto dense cores, and within toward protostars. Most known streamers have been discovered serendipitously \citep[e.g.,][]{LeGouellec2019, Pineda2020, Murillo2022, Valdivia-Mena2022}, often revealed through kinematic studies in molecular line tracers. Although kinematic information is necessary to confirm the infalling nature of streamers, the morphology of the dust continuum can provide tentative clues to their presence. Here, we highlight several sources in our sample that exhibit streamer-like morphology in the 12~m emission, which would be good candidates to follow-up with higher angular resolution spectral line data.

\begin{figure*}
    \centering
        \gridline{
        \fig{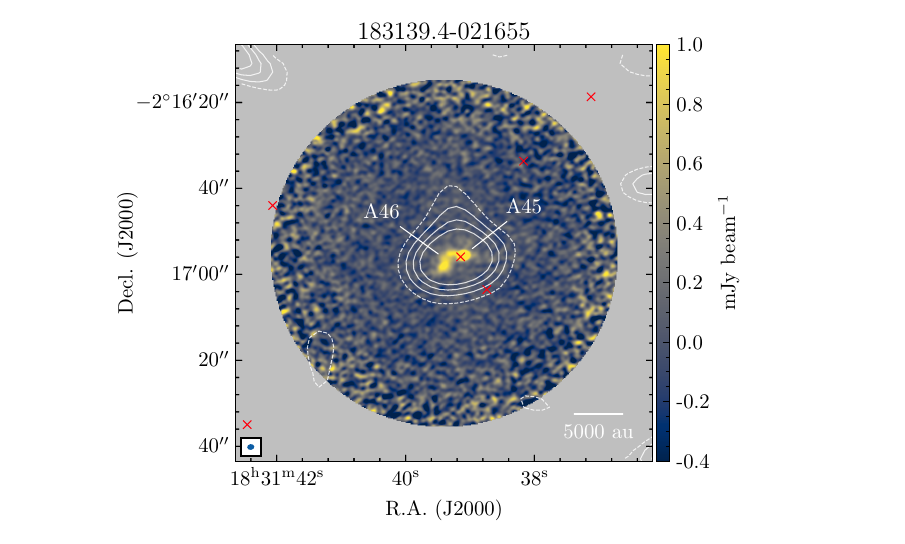}{0.85\textwidth}{}
    }
    \gridline{
        \fig{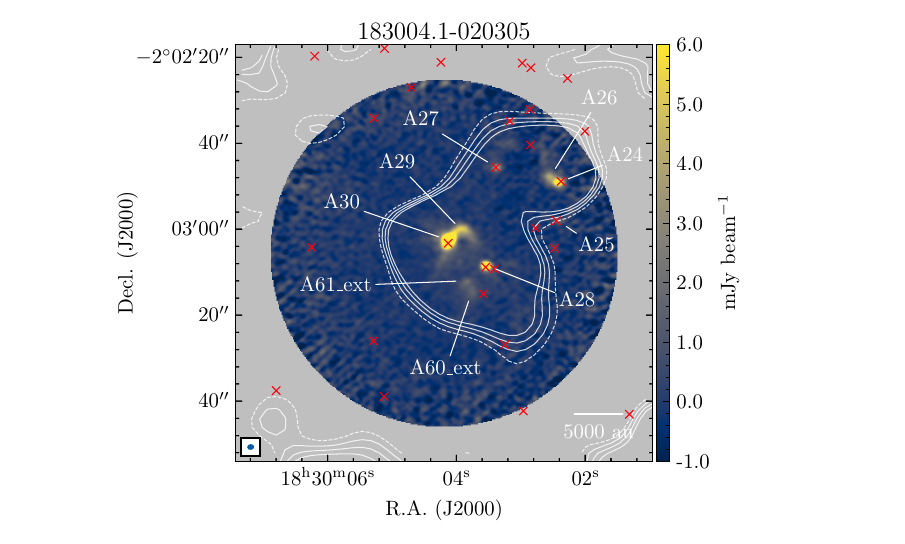}{0.85\textwidth}{}
    }
    \caption{
    ALMA 12~m emission showing potential streamer-like morphologies.
    Top panel: ALMA 12~m field 183139.4-021655 containing the starless candidate source A46. Bottom panel: ALMA 12~m field 183004.1-020305 containing the starless candidate sources A26, A29, A60\_ext, and A61\_ext. See Figure \ref{field-020349Mosaic-ref} for plotting conventions.
    \label{fig:possible-streamer-candidates}
    }
\end{figure*}

The top panel of Figure \ref{fig:possible-streamer-candidates} shows ALMA 12~m sources A45 (protostellar) and A46 (candidate starless), separated by a distance of approximately \ang{;;4.5}. There appears to be an arc of dust continuum emission found between the protostar and the candidate starless core, toward the northeast. Further kinematic observations could serve to identify the nature of this emission and whether or not it is infalling to both (or either) structures. If indeed, source A46 contains no deeply embedded protostellar sources, this substructure may be in a state where it is feeding the nearby protostar and envelope structure, analogous to a proposed scenario for the Per-emb-2 system \citep{Taniguchi2024}.

Additionally, the bottom panel of Figure \ref{fig:possible-streamer-candidates} shows HGBS dense core 183004.1-020305 containing source A30 at its center. This detection is accompanied by an arc of dust emission (identified as A29) to the northwest. These detections, along with the many other substructures found in the nearby region, comprise the core with the most amount of fragmentation in our observed sample. \citet{Plunkett2018} performed an ALMA Band 6 (1~mm) continuum study with slightly higher angular resolution (\ang{;;1.0}) of this region, and identified many unstable substructures in the same region of arc-like emission, which they label as the central ``ridge."

\subsection{Group A and B Protostellar Sources: Undetected and ACA-only Detections}
\label{sec:appendix-protostellar-only-7m}

Generally, protostars are detected easily in our ALMA 12~m observations; however, there were 11 instances where protostellar HGBS cores were not detected in our sample. Of these, four were in Group B (detected only in the 7~m data), and seven were in Group A (no detections in either the 12~m or 7~m data). We examine these sources in detail here to better understand the nondetections and supply an overview of these details in Table \ref{tab:undetected-cores}.

\begin{deluxetable}{lccccccc}
\tablecaption{
Properties of undetected protostellar dense cores. \label{tab:undetected-cores}
}
\tablehead{
    \colhead{HGBS Source} &
    \colhead{Herschel 70~$\mu$m\tnm{a}} &
    \multicolumn{3}{c}{Protostar Association\tnm{b}} &
    \colhead{7~m Det.} &
    \colhead{12~m Det.\tnm{c}} &
    \colhead{SCUBA 450~$\mu$m\tnm{d}} \\[-0.5pt]
    \colhead{Name} &
    \colhead{} &
    \colhead{Sep. (arcsec)} &
    \colhead{Class} &
    \colhead{Catalog} &
    \colhead{} &
    \colhead{($<3\sigma$)} &
    \colhead{}
}
\startdata
\sidehead{Group A Cores}
182908.3-020528 & N & 9.88 & II & (1) & N & N & Y \\
182957.5-015843 & N & 7.25 & II & (3) & N & N & Y \\
183016.2-020716 & Y & 9.4\tnm{*} & 0 & (1) & N & N & Y \\
183018.1-020843 & N & 2.74 & II & (3) & N & N & Y \\
183102.6-021016 & N & 8.18 & 0 & (5) & N & N & Emerging \\
183157.4-020025 & N & 4.27 & 0/I & (5) & N & N & Emerging \\
183236.3-014841 & N & 5.86 & II & (4) & N & N & N \\
\sidehead{Group B Cores}
183002.7-020103 & N & 8.16 & 0 & (5) & Y & N & Emerging \\
183014.9-013334 & Y & 6.75 & 0 & (2) & Y & Y & Emerging \\
183148.3-020139 & N & 7.70 & 0 & (5) & Y & Y & Y \\
183229.0-015242 & N & 3.54 & I & (1) & Y & N & Emerging \\
\enddata

\tablenotetext{a}{Sources marked ``Y" are those which show a compact 70$\mu$m source in the original HGBS catalog (i.e., classified in the HGBS catalog as protostellar). All sources marked ``N" are those we reclassified as protostellar based on catalog associations (see Section \ref{sec:protostellar-associations} for details).}

\tablenotetext{b}{We show the separation, YSO class, and YSO catalog of the closest protostar to the center of the HGBS dense core. The following is a list of the catalogs used and their entry in this table: (1) SESNA \citep[R. Gutermuth et al., 2026, in preperation;][]{Pokhrel2020}; (2) eHOPS \citep{Pokhrel2023}; (3) MystIX \citep{Povich2013}; (4) W40 \citep{Mallick2013}; (5) the multiwavelength catalog from \citet{Maury2011}.}

\tablenotetext{c}{Sources marked ``Y" are starting to emerge in the ALMA 12~m data, with emission levels below a threshold of $3\sigma$.}

\tablenotetext{d}{Sources marked ``Y" are those which show clear and compact SCUBA 450~$\mu$m emission. Source marked ``Emerging" show clear but extended emission.}

\tablenotetext{*}{This SESNA Class 0 YSO lies \textit{outside} of the \ang{;;6.5} deconvolved core radius of HGBS dense core 183016.2-020716. While this would mean it falls outside the association criteria (as outlined in Section \ref{sec:protostellar-reclassification}), the HGBS dense core was originally classified as protostellar due to the compact 70~$\mu$m source.}

\end{deluxetable}

\begin{figure*}
    \centering
    \noindent
    \includegraphics[width=0.5\textwidth]{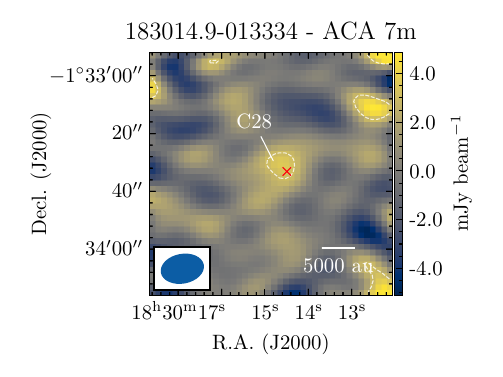}%
    \includegraphics[width=0.5\textwidth]{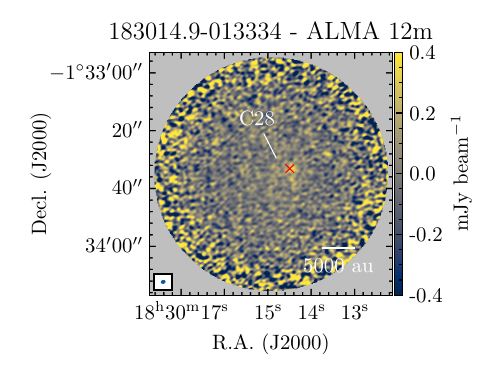}\\[0pt]
    \includegraphics[width=0.5\textwidth]{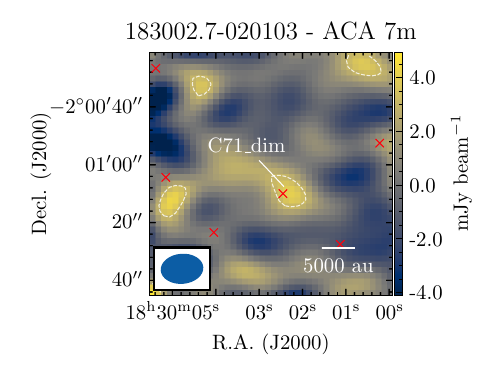}%
    \includegraphics[width=0.5\textwidth]{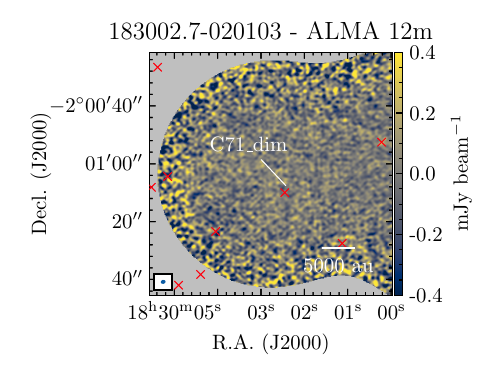}\\[0pt]
    \includegraphics[width=0.5\textwidth]{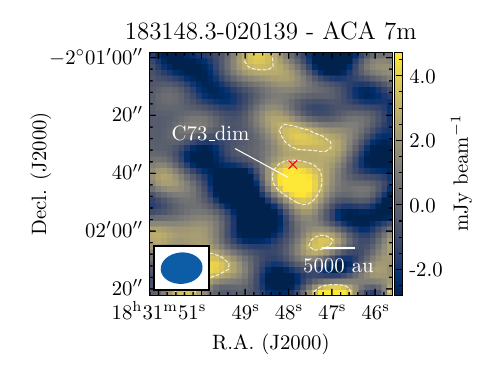}%
    \includegraphics[width=0.5\textwidth]{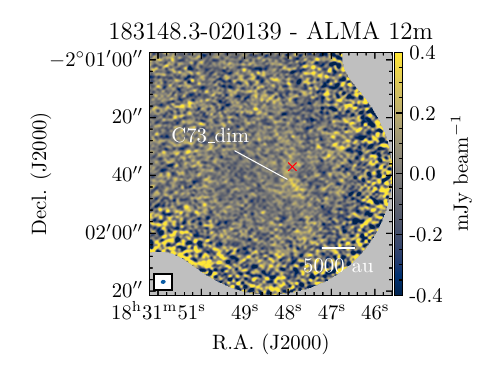}
    \caption{
    HGBS protostellar dense cores only detected in ACA 7~m emission. Panel titles indicate the HGBS dense core and associated ALMA dataset. Plotting conventions generally follow those found in Figure \ref{field-020349Mosaic-ref}.
    \label{fig:7m-only-1}
    }
\end{figure*}

\begin{figure*}
    \centering
    \noindent
    \includegraphics[width=0.5\textwidth]{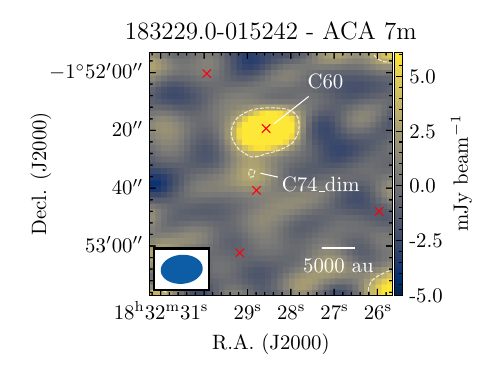}%
    \includegraphics[width=0.5\textwidth]{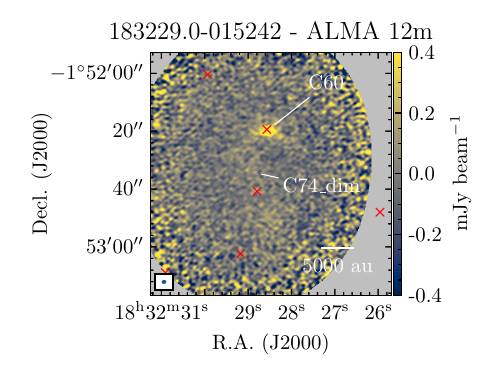}
    \caption{
    HGBS protostellar dense cores only detected in ACA 7~m emission. Panel titles indicate the HGBS dense core and associated ALMA dataset. Plotting conventions generally follow those found in Figure \ref{field-020349Mosaic-ref}.
    \label{fig:7m-only-2}
    }
\end{figure*}

In the four Group B cores, we detect relatively faint emission in the ACA 7~m data, with no corresponding robust ALMA 12~m emission at the same locations classified as protostellar. These examples are shown in Figures \ref{fig:7m-only-1} and \ref{fig:7m-only-2}, where the ACA 7~m detections are shown on the left (along with a $3\sigma$ contour and the protostar locations), and the corresponding 12~m emission map is shown on the right. Within two of these 7~m detections, sources C28 and C73\_dim, we tentatively detect 12~m continuum emission slightly below the $3\sigma$ threshold. In addition, our CO data suggests a bipolar outflow originating from the location of C28, confirming the protostellar nature of this detection. For these Group B cores, the 7~m emission is weak, and very likely extended on angular scales approaching the 12~m array's largest recoverable scale of $\sim\ang{;;18}$, suggesting that the corresponding emission is resolved out at higher angular resolution and, hence, its nondetections in the 12~m data.

Considering both groups together, the separations between the HGBS core centers and their associated protostars range from approximately \ang{;;3}-\ang{;;10} (see Table \ref{tab:undetected-cores}), placing the associated protostars toward the outskirts of the dense core rather than near its center. These larger separations raise the possibility that some of these associations are chance alignments rather than physically connected sources, a point we return to when discussing the Group A cores below.

We retain the protostellar classification in these cases to remain conservative regarding the identification of starless detections. It is nonetheless peculiar that the associated sources are all classified as Class 0/I YSOs yet show no detectable (or only emerging) compact 12~m emission. If these associations are genuine, the nondetections imply that the compact dust and gas reservoirs surrounding the protostar are below the $3\sigma$ sensitivity limit of the 12~m observations, 0.3~mJy~beam$^{-1}$, corresponding to a mass of $\sim0.045M_{\odot}$ using the assumptions in Section \ref{sec:derived_properties}, suggesting that these may be very low-mass systems.

In addition to the protostellar dense cores exclusively detected with the ACA 7~m array, there are seven protostellar dense cores that remain undetected in both the ALMA 12~m and ACA 7~m data. Four of these are associated with Class II YSOs, with the remaining three classified as Class 0/I sources. We note that four of these associations (out of the total seven in Group A) are on the larger end of our matching criteria (see Section \ref{sec:protostellar-associations} and Table \ref{tab:undetected-cores}), with separations between the HGBS core center and protostar location approaching the deconvolved core radius, and should therefore be treated with some caution. Interestingly, three of the four Class II associated cores show clear emission in the SCUBA-2 450~$\mu$m data at their locations, suggesting that dust emission is present but falls below the sensitivity threshold for our ALMA observations. In general, the nondetection of even the Class 0/I sources is somewhat surprising, as the dusty envelope surrounding a centrally embedded protostar would be expected to produce detectable continuum emission at these wavelengths.

There are two possible explanations for these nondetections in both the ACA 7~m data and the ALMA 12~m data: (1) the sources are genuinely fainter than the sensitivity limits of both arrays, or (2) they have been misclassified.
In the former case, the $3\sigma$ sensitivity of the 7~m array of 3.75~mJy~beam$^{-1}$ implies a mass upper limit of $\sim0.56M_{\odot}$ on scales smaller than approximately \ang{;;60}. As stated above, the 3$\sigma$ sensitivity limit of the 12~m observations corresponds to an upper mass limit of $\sim0.045M_{\odot}$ on scales smaller than approximately \ang{;;18}.

We now briefly consider the possibility of contamination in the protostellar catalog, e.g., from extragalactic sources, is a possible source of mis-classification. The SESNA catalog estimates $9\pm1$ residual extragalactic contaminants per square degree, distributed roughly equally between Class I and Class II SED classes \citep[R. Gutermuth et al. 2026, in preperation;][]{Pokhrel2020}, suggesting that the contamination is not preferentially biased toward any particular evolutionary stage.
Taking the primary beam of $\sim$~\ang{;;82}, corresponding to an area of $\sim4\times10^{-4}$ square degrees per pointing, the total ALMA coverage across 100 fields spans approximately 0.04 square degrees. At the SESNA estimated contamination rate of $9\pm1$ extragalactic sources per square degree, we would expect fewer than one (0.4) contaminant across the full survey area. It therefore seems unlikely that extragalactic contamination can account for all seven nondetections in Group A.

Further insight could be obtained through molecular line observations. Genuine Class 0/I YSOs would be expected to retain significant reservoirs of dense gas tracers such as N$_2$H$^{+}$ or NH$_3$, signatures that would be absent in contaminants such as extragalactic sources.

\section{ALMA 12~m $^{12}$CO$(1-0)$ Observations}
\label{sec:appendix-12co}

As stated in Section \ref{sec:alma-co-data}, there exists many instances of detectable outflow structure in our 12~m CO observations. We show four 12~m continuum detections in three dense cores that are labeled as ``protostellar" by \citet{Konyves2015}, but have not been found to be protostellar by any other protostellar catalogs that we examined. Three of these four 12~m continuum detections are toward the east, which have minimal protostellar catalog overlap, as mentioned in Section \ref{sec:alma-co-data}.

\begin{figure*}
    \centering
    \noindent
    \includegraphics[width=0.5\textwidth]{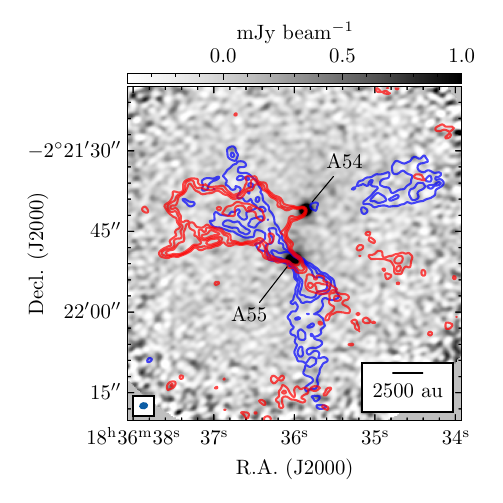}%
    \includegraphics[width=0.5\textwidth]{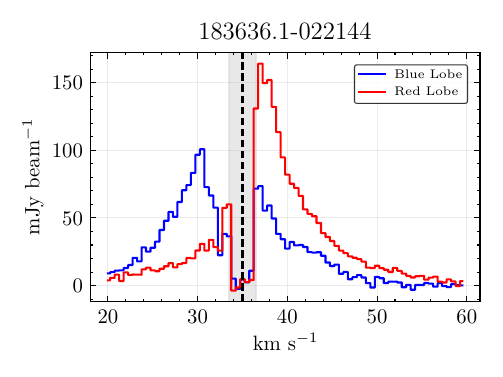}\\[0pt]
    \includegraphics[width=0.5\textwidth]{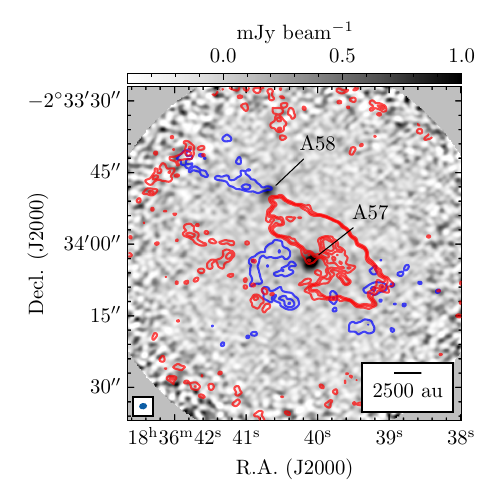}%
    \includegraphics[width=0.5\textwidth]{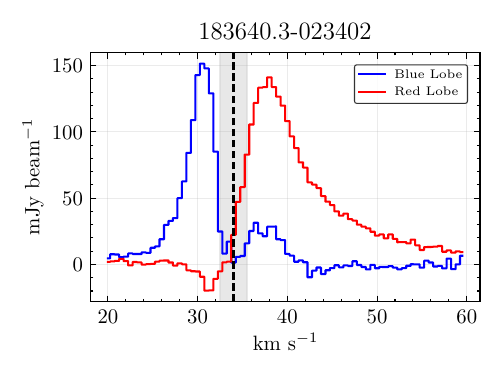}
    \caption{
        CO outflow signatures for ALMA 12~m detections without direct protostellar associations. Left panels: ALMA 12~m continuum emission is shown in grayscale ranging linearly from $-0.4-1.0$~mJy~beam$^{-1}$. The synthesized beam is given in the bottom left corner and a scalebar in the bottom right corner. The blue and red contours represent the velocity-shifted components of the $^{12}$CO$(2-1)$ data, with integrated velocity ranges on either side of the central velocity region (see below). The contours are drawn at $3\sigma$, $5\sigma$, $7\sigma$, where the noise is taken from a representative noise-free channel. Right panels: spatially averaged spectrum of the 3$\sigma$ contours of both sets of contours on the left. The gray shading indicates the velocity range excluded from the lobes, and the vertical dashed black line indicates the midpoint of this range. We give the ALMA field name for which the accompanying sources are found above the plot.
        \label{fig:source55-source57-co}
        }
\end{figure*}

The top panel of Figure \ref{fig:source55-source57-co} shows field 183636.1-022144 with detections A54 and A55. Both sources contain strong outflow emission at approximately perpendicular angles. Source A54, the upper source, shows blue shifted material ejected to the west, while there is strong red-shifted material emission to the east. Source A55 on the other hand seems to suggest two outflows at the same angle. There is significant blue- and red-shifted emission to the southwest and northeast.

The bottom panel of Figure \ref{fig:source55-source57-co} shows field 183640.3-023402 with sources A57 and A58, in which source A58 has been classified as protostellar through the SPICY catalog. Source A58 shows very strong red-shifted outflow detection to the southwest with a relatively small blue-shifted outflow detection to the northeast. Source A57 shows a relatively weaker outflow structure, with an arc of blue-shifted material to the southeast, and red-shifted emission (although made difficult to parse with the A58 red-shifted outflow) to the northwest.

\begin{figure*}
    \centering
    \noindent
    \includegraphics[width=0.5\textwidth]{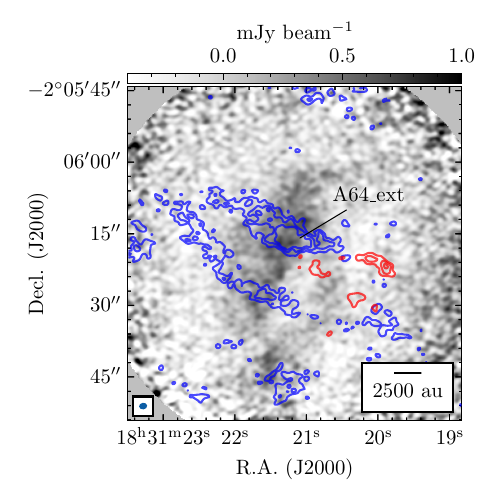}%
    \includegraphics[width=0.5\textwidth]{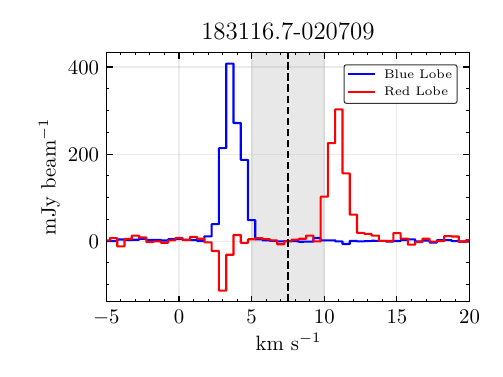}
    \caption{
        Left panel: zoomed-in view of the ALMA field 183116.7-020709Mosaic containing source A64\_ext. The integrated velocity ranges of the blue- and red-shifted components are $-30-5.0$~km~s$^{-1}$ and $10.0-29.5$~km~s$^{-1}$, respectively. The contours are drawn at $3\sigma$, $5\sigma$, $7\sigma$, where the noise is $1\sigma\sim$~0.135 mJy~beam$^{-1}$. See Figure \ref{fig:source55-source57-co} for all other plotting conventions.
    \label{fig:source64_ext-co}
    }
\end{figure*}

Finally, Figure \ref{fig:source64_ext-co} shows extended continuum detection A64\_ext, found within the protostellar dense core 183121.1-020619. As stated in Section \ref{sec:alma-co-data}, there are no protostellar associations with the catalogs we searched other than the initial \citet{Konyves2015} classification, which is guided by a compact source found within the 70~$\mu$m Herschel data. We find some red shifted emission to the west of the location of the ALMA 12~m continuum emission, and some larger and extended blue emission structures centered on the 12~m detection and to the south. Since the CO emission has an ambiguous morphology and could arise from an extended CO structure that is partially resolved out, we classify A64\_ext as starless. Our analysis in Section \ref{sec:predicted-detection-count} concerns only detections within starless HGBS dense cores; thus, the classification of this source does not affect our statistics regardless of the interpretation of the CO data.

\section{Bonnor-Ebert Spheres}
\label{sec:appendix-be-spheres}

To construct the BE sphere models, we utilize the \citet{Dapp2009} analytic function to approximate a BE-sphere density profile, assuming six different central number densities, ranging from $10^5-10^{10}$~cm$^{-3}$. The input parameters include a temperature of $T=10$~K, along with a truncation radius of 6800~au, the mean observed core radii in the prestellar population of the HGBS core catalog \citep{Konyves2015}. We then appropriately model the temperature when computing the flux maps in order to be consistent with the temperature modeling used for the simulations. These are the same temperature models as shown in Figure 16 of \citet{Dunham2016}.

\begin{figure*}
    \includegraphics[]{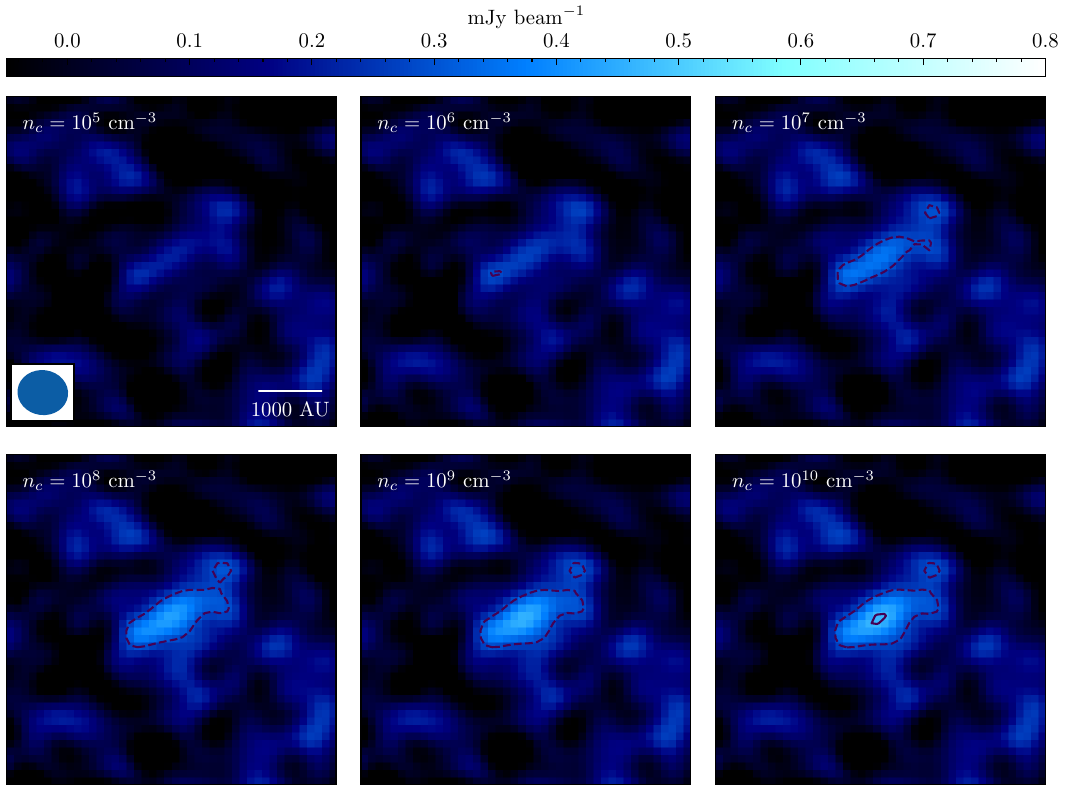}
    \caption{
    Synthetic ALMA 106~GHz observations of the constructed BE-sphere models, given at six central number densities. Each panel indicates the central number density. The synthesized beam and scale bar are given in the first panel. Dashed contours represent the $3\sigma$ level of emission, while solid contours represent the $5\sigma$ level and increase by $2\sigma$, where $1\sigma$ rms is $\sim$~0.09~mJy~beam$^{-1}$.
    \label{fig:BE-sphere-simulations}
    }
\end{figure*}

Figure \ref{fig:BE-sphere-simulations} shows the results from the synthetic observations where the core is undetectable to $5\sigma$ until the central number density reaches approximately $10^{10}$~cm$^{-3}$. We also test the changes in detectability for alternative choices of truncation radius, specifically factors of 2 less ($r=3400$~au) and 2 more ($r=13600$~au). The truncation radius has a minimal effect on the inner flat region; the inner flat region is primarily controlled by the central number density. We find no changes in detectability in these two alternative cases.

\section{Supplementary Figures -- ALMA 12~m Detections}
\label{sec:appendix-extra-12m-figures}

In this section, we provide a dedicated figure set that contains all other ALMA 12~m continuum images that yielded positive detections and that have not been shown in the text previously. Fields are mosaicked together when possible for better sensitivity (see Section \ref{sec:calibration-reduction}) and follow the conventions as outlined in Figure \ref{field-020349Mosaic-ref}.

\figsetstart
\figsetnum{20}
\figsettitle{Supplementary ALMA 12m Detections}

\figsetgrpstart
\figsetgrpnum{20.23}
\figsetgrptitle{183116.7-020709Mosaic}
\figsetplot{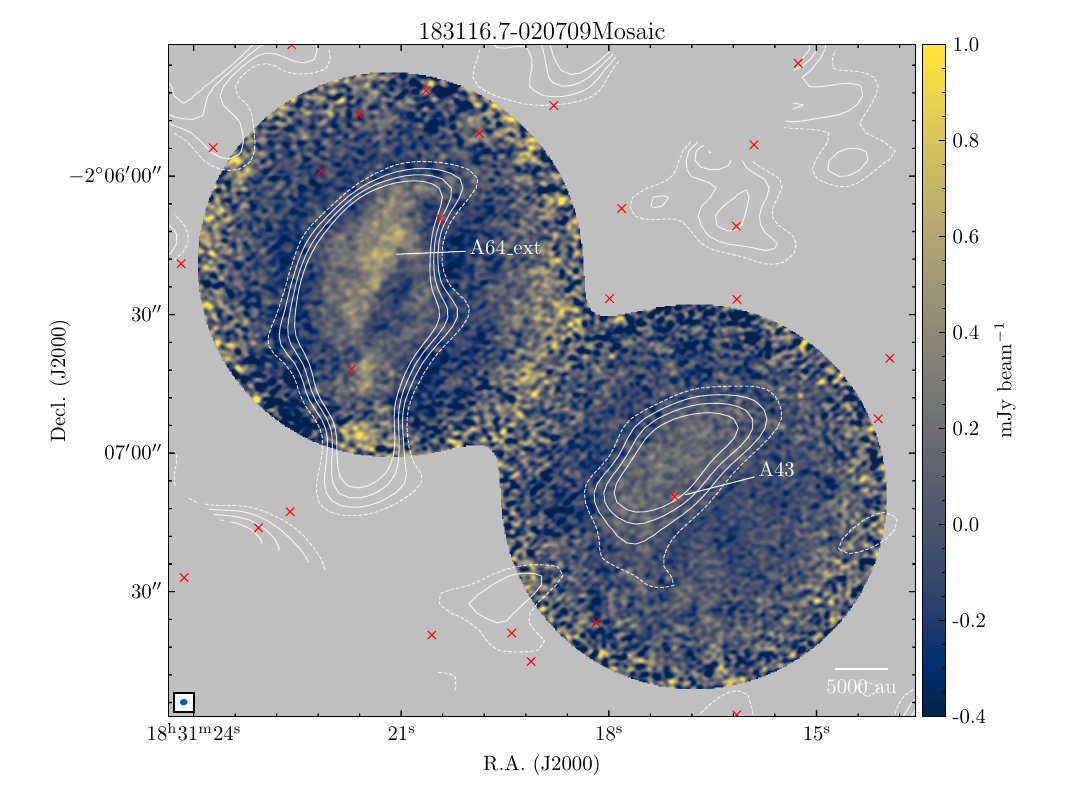}
\figsetgrpnote{ALMA 12m field 183116.7-020709Mosaic containing continuum detection(s). See Figure 2 for plotting conventions.}
\figsetgrpend

\figsetgrpstart
\figsetgrpnum{20.24}
\figsetgrptitle{183148.3-020139Mosaic}
\figsetplot{FigureAppE_24_183148.3-020139Mosaic_12m_ALMA_Detections.pdf}
\figsetgrpnote{ALMA 12m field 183148.3-020139Mosaic containing continuum detection(s). See Figure 2 for plotting conventions.}
\figsetgrpend

\figsetgrpstart
\figsetgrpnum{20.25}
\figsetgrptitle{183228.8-015221Mosaic}
\figsetplot{FigureAppE_25_183228.8-015221Mosaic_12m_ALMA_Detections.pdf}
\figsetgrpnote{ALMA 12m field 183228.8-015221Mosaic containing continuum detection(s). See Figure 2 for plotting conventions.}
\figsetgrpend

\figsetend

\begin{figure*}
    \digitalasset
    \centering
    \includegraphics[]{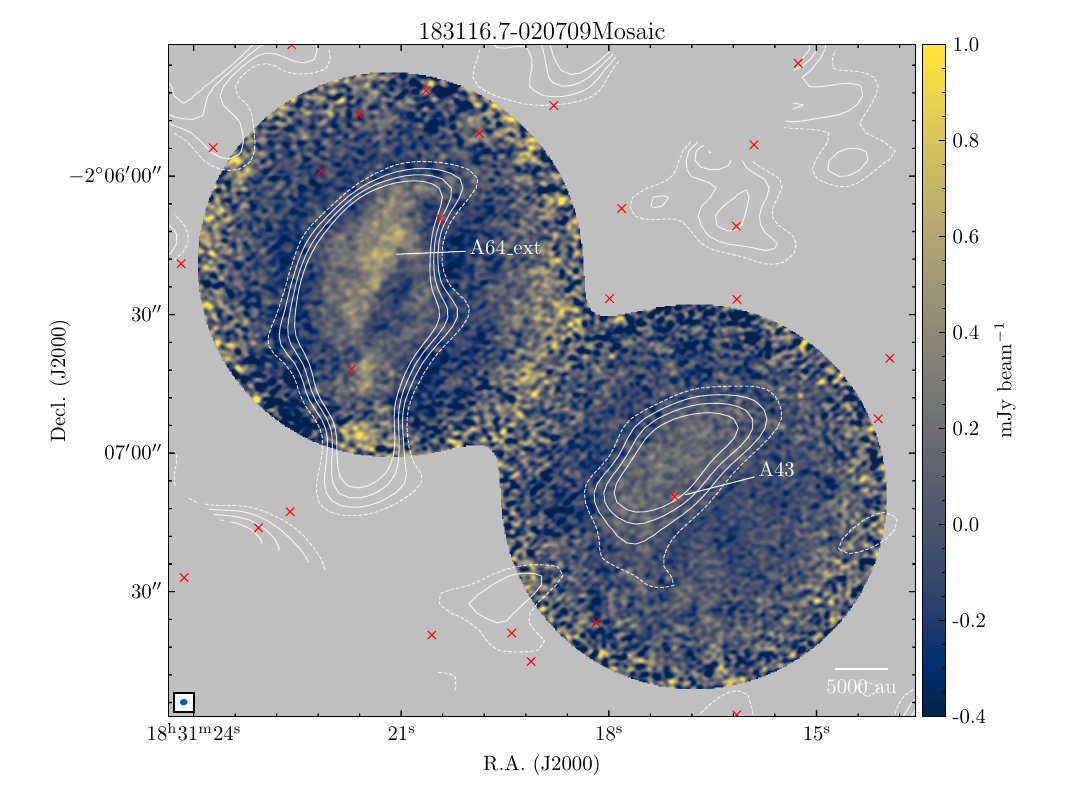}
    \caption{
    All ALMA 12~m continuum detections not previously shown in the text.
    ALMA 12~m field 183116.7-020709Mosaic containing continuum detection(s). See Figure \ref{field-020349Mosaic-ref} for plotting conventions. \\ \\
    (The complete figure set (25 images) is available in the online journal.)
    \label{fig:alma-detections-figureset}
    }
\end{figure*}

\section{Supplementary Figures -- Multi-scale Observations}
\label{sec:appendix-supplementary-multiscale}

In this section, we present all observed dense cores in a four-panel format containing: (1) the Herschel column density map; (2) the SCUBA-2 450$\mu$m flux map; (3) the ALMA-ACA 7~m observations; and (4) the ALMA 12~m observations. We adopt the ALMA-ACA 7~m observations as the reference for these figures; we use the mosaics (as labeled in Table \ref{tab:observation-noise-levels}) when we are able, and revert to the individual fields as necessary.

\figsetstart
\figsetnum{21}
\figsettitle{Four-Panel Multiscale Dataset Figures}

\figsetgrpstart
\figsetgrpnum{21.14}
\figsetgrptitle{182908.3-013046}
\figsetplot{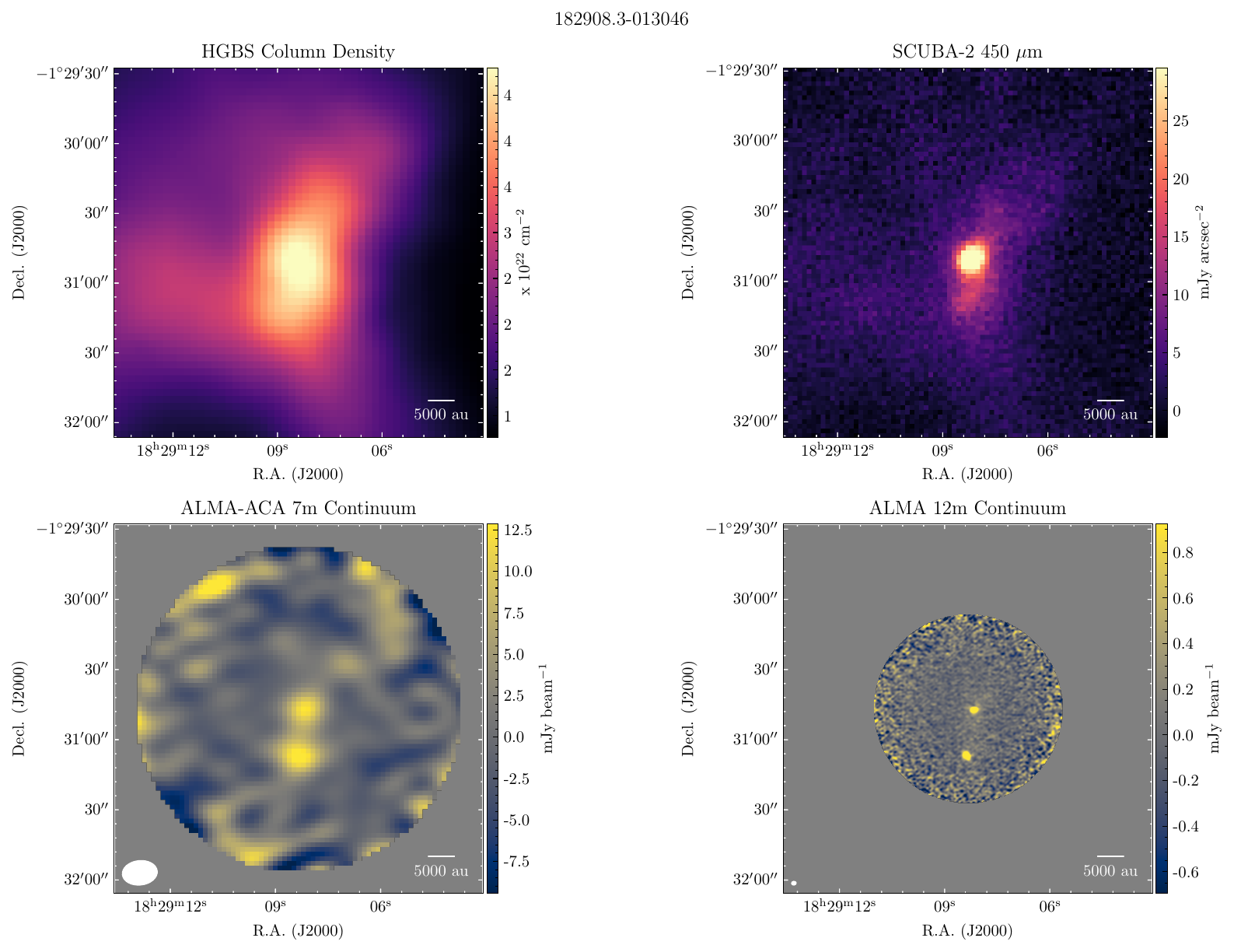}
\figsetgrpnote{A four-panel view of the observed dense core(s), centered on the ACA 7m field 182908.3-013046. The top left panel contains the Herschel H$_2$ column density map, the top right panel contains the SCUBA-2 450~$\mu$m emission map, the lower left panel contains ACA 7m continuum emission map, and the bottom right panel contains the ALMA 12m continuum emission map. All panels are linearly normalized to their respective 99th percentile range; colorbars are displayed for each dataset. We display a scale bar on each panel of 5000~au and plot the beam size in the lower two panels for the ALMA-based datasets.}
\figsetgrpend

\figsetend

\begin{figure*}
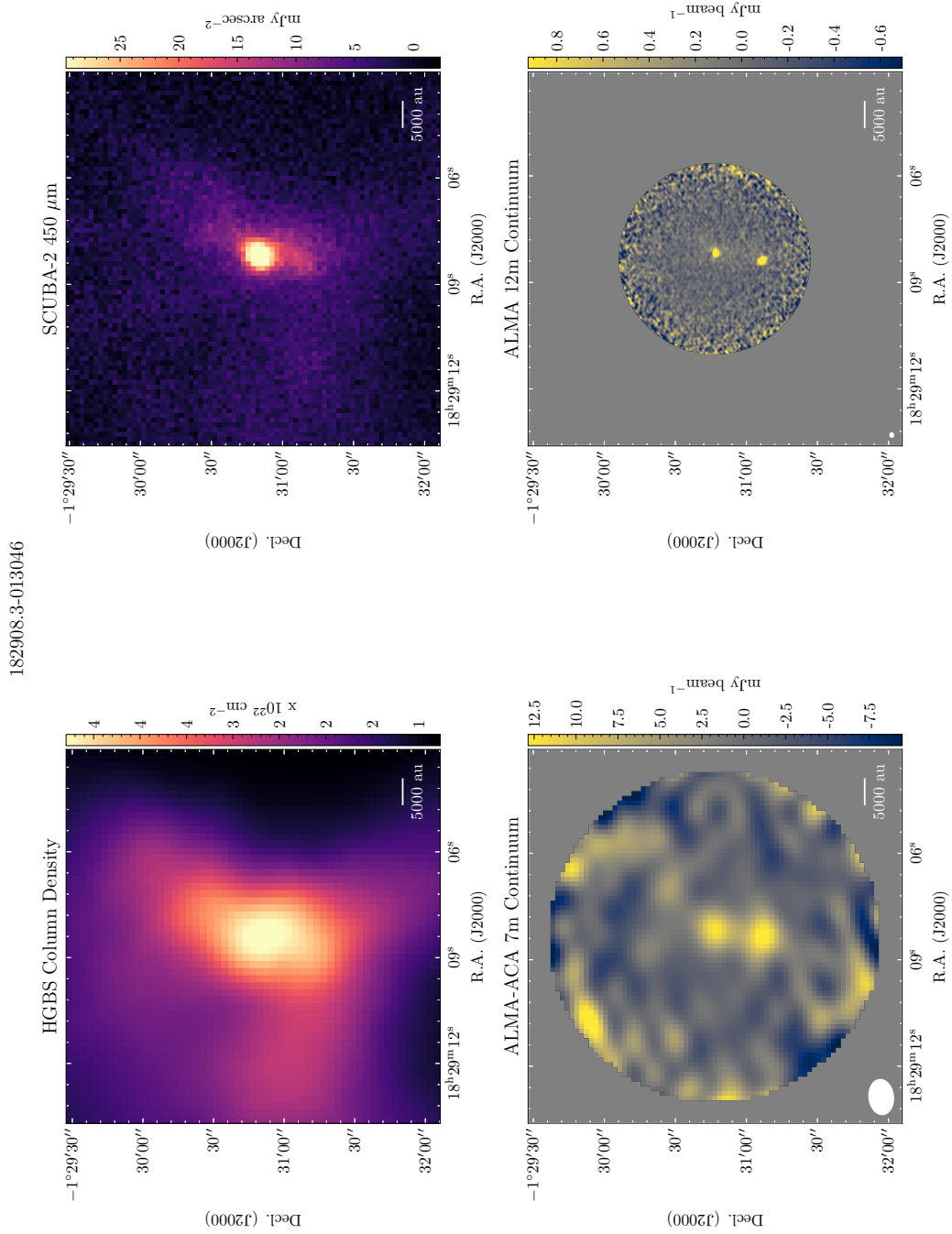

    \digitalasset
    \centering
    \rotatefig{90}{FigureAppF_13_182908.3-013046.pdf}{\textwidth}{}
    \caption{
    A four-panel view of the observed dense core(s), centered on the ACA 7~m field 182908.3-013046. The top-left panel contains the Herschel H$_2$ column density map, the top-right panel contains the SCUBA-2 450~$\mu$m emission map, the bottom-left panel contains ACA 7~m continuum emission map, and the bottom-right panel contains the ALMA 12~m continuum emission map. All panels are linearly normalized to their respective 99th percentile range; a color bar is displayed for each dataset. We display a scale bar on each panel of 5000~au and plot the beam size in the bottom two panels for the ALMA-based datasets. \\ \\
    (The complete figure set (60 images) is available in the online journal.)
    \label{fig:multiscale-figureset}
    }
\end{figure*}

\clearpage
\bibliography{references_aquila_2026}{}
\bibliographystyle{aasjournal}

\end{document}